\documentclass[journal]{IEEEtran}
\ifCLASSINFOpdf

\else

\fi
\usepackage{mathrsfs}

\usepackage{amsmath}
\usepackage{ntheorem}

\usepackage{makeidx}  
\usepackage{algorithm}
\usepackage{algorithmic}
\usepackage{graphicx}
\usepackage{subfigure}
\usepackage{epstopdf}
\usepackage{bm}
\usepackage{cite}
\usepackage{stfloats}
\usepackage{setspace}

\usepackage{amssymb}
\usepackage{graphicx}
\newtheorem{theorem}{Theorem}
\usepackage{url}

\usepackage{tabularx,booktabs}
\newcolumntype{C}{>{\centering\arraybackslash}X} 
\usepackage{lipsum}

\usepackage{makecell} 

\usepackage{graphicx}
\usepackage{color}

\newtheorem{lem}{Lemma}

\newtheorem{definition}{Definition}
\usepackage{multirow}

\definecolor{dblue}{RGB}{15,89,164}

\begin{document}

\title {Performance Analysis and Low-Complexity Design for XL-MIMO with Near-Field Spatial Non-Stationarities}
\author{	Kangda Zhi, Cunhua Pan, Hong Ren, Kok Keong Chai, Cheng-Xiang Wang, \IEEEmembership{Fellow, IEEE},  Robert Schober, \IEEEmembership{Fellow, IEEE}, Xiaohu You, \IEEEmembership{Fellow, IEEE}\thanks{
		 Part of this work has been accepted by the EUSIPCO, Helsinki, Finland, Sept. 4-8, 2023 \cite{zhi2023euspcio}. \itshape(Corresponding author: Cunhua Pan.)\upshape 
		
		The work of Cunhua Pan was supported in part by the National Natural Science Foundation of China (Grant No. 62201137,62331023). The work of Cunhua Pan and Hong Ren was supported in part by the Fundamental Research Funds for the Central Universities, 2242022k60001, and  by the Research Fund of National Mobile Communications Research Laboratory, Southeast University (No.2023A03). The work of H. Ren was supported in part by the National Natural Science Foundation of China (Grant No. 62101128), and in part by Basic Research Project of Jiangsu Provincial Department of Science and Technology (Grant No. BK20210205). The work of Kangda Zhi was supported by China Scholarship Council. The work of C.-X. Wang was supported by the National Natural Science Foundation of China (NSFC) under Grant 61960206006, the Fundamental Research Funds for the Central Universities under Grant 2242022k60006, the Key Technologies R$\&$D Program of Jiangsu (Prospective and Key Technologies for Industry) under Grants BE2022067 and BE2022067-1, and the EU H2020 RISE TESTBED2 project under Grant 872172.
		 		
		K. Zhi and K. K. Chai are with the School of Electronic Engineering and Computer Science at Queen Mary University of London, UK. (e-mail: k.zhi, michael.chai@qmul.ac.uk).
		
		C. Pan, H. Ren, C.-X. Wang, and X. You are with the National Mobile Communications Research Laboratory, Southeast University, China. (cpan, hren, chxwang, xhyu@seu.edu.cn).
		
		R. Schober is with the Institute for Digital Communications, Friedrich-Alexander-University Erlangen-N\"{u}rnberg (FAU), Germany (e-mail: robert.schober@fau.de).

}
}

\maketitle

\begin{abstract}
Extremely large-scale multiple-input multiple-output (XL-MIMO) is capable of supporting extremely high system capacities with large numbers of users. In this work, we build a framework for the analysis and low-complexity design of XL-MIMO in the near field with spatial non-stationarities. Specifically,  we first analyze the theoretical performance of  discrete-aperture XL-MIMO using an electromagnetic (EM) channel model based on the near-field spherical wavefront. We analytically  reveal the impact of the discrete aperture and polarization mismatch on the received power. We also complement the classical Fraunhofer distance based on the considered EM channel model. Our analytical results indicate that a limited part of the XL-array receives the majority of the signal power in the near field, which leads to a notion of visibility region (VR) of a user. Thus, we propose a  VR detection algorithm and leverage the acquired VR information to devise a low-complexity symbol detection scheme. Furthermore, we propose a graph theory-based user partition algorithm, relying on the VR overlap ratio between different users. Partial zero-forcing (PZF) is utilized to eliminate only the interference from users allocated to the same group, which further reduces computational complexity in matrix inversion. Numerical results confirm the correctness of the  analytical results and the effectiveness of the proposed algorithms. It reveals that our algorithms approach the performance of conventional whole array (WA)-based designs but with much lower complexity. 
\end{abstract}


\IEEEpeerreviewmaketitle

\section{Introduction}
To fulfill the various challenging demands on the fifth-generation (5G) and beyond wireless systems,  several appealing technologies have been proposed and investigated, including massive multiple-input multiple-output (MIMO) \cite{bjornson2017massive,bjornson2015massive,zhang2020antenna,zhi2022ZF}, cell-free \cite{bjornson2020scalable}, small cell \cite{chen2018stochastic}, millimeter-wave (mmWave) \cite{Heath2016mmwave}, and reconfigurable intelligent surface (RIS) \cite{pan2020intelligent,huang2019reconfigu,pan2020multicell,chen2022active}. Among them, as the evolution of massive MIMO, extremely large-scale MIMO (XL-MIMO) has recently garnered new interest\cite{bjornson2019massive,wang2022extremely}. By mounting several thousands of antennas, XL-MIMO can achieve extremely high spectral efficiency and fulfill the harsh data rate requirements of  future wireless systems. XL-MIMO may have large physical dimensions spanning several tens of meters\cite{de2020non}. It is expected to be integrated into large structures such as the walls of buildings in mega-city, airports, large shopping malls, and stadiums, enabling simultaneous service to a significant number of users.

A classic criterion for distinguishing the boundary between the near and far fields is the Fraunhofer (Rayleigh) distance $d_f=\frac{2D^2}{\lambda}$, where $D$ and $\lambda$ denote the largest array aperture and wavelength, respectively\cite{selvan2017fraunhofer}. As  $D$ increases, boundary $d_f$ expands, and the users will be easily located in the near field of the XL-MIMO instead of the far field. Accordingly, the practical spherical electromagnetic (EM) wavefront can no longer be approximated as planar wavefront.  There are several distinctions  between near-field and far-field communications. The first one is the nonlinear variation of the phase of the received signal across the whole array. Under the far-field condition, the phase of the array steering vector is approximately linearly scaled for different elements, which makes mathematical analysis tractable. However, this characteristic does not hold in the near field. Secondly, as the array aperture increases, it is essential to consider the amplitude/power variation across the entire array. This is because the distance between the user and the array center could be significantly different from that between the user and the array edge. Thirdly, in the near field, the incident wave directions received by distinct antenna elements within the array may exhibit considerable disparities, resulting in substantially varying projected apertures across the whole array. Therefore, for studying XL-MIMO, it is crucial to consider the practical spherical wavefront and  investigate the new features introduced by  near-field communications.

Taking into consideration  the near-field behavior with respect to the whole array, XL-MIMO has recently been studied  from different perspectives. Focusing on the nonlinear phases of the array steering vector, some work has investigated beam training\cite{cui2022near,liu2022deep,You2022NearTrain}, channel estimation \cite{cui2022channel}, and multiple-access design\cite{wu2022multiple}. To further accurately model the near-field spherical wavefront, the variation of the amplitude across the array was considered in \cite{hu2018beyond,Dardari2020LIS,bijoson2020nearField,de2020near,bjornson2021primer,pizzo2022holo,wei2023channel,dardari2021access}. Specifically, the authors of \cite{hu2018beyond} modeled the near-field channel accounting for the variations of  the amplitude and incident wave direction across the whole array. Derived from Maxwell's equations, the authors of  \cite{Dardari2020LIS,bijoson2020nearField,de2020near} adopted an EM channel model and characterized the EM polarization effect, which accurately depicts the physical near-field behavior.  These works proved that due to the amplitude attenuation at the array edge, despite the aperture of the array approaching infinity, the received power of the signal remains limited.  However, for tractability, the above contributions \cite{hu2018beyond,Dardari2020LIS,bijoson2020nearField,de2020near,bjornson2021primer,pizzo2022holo,wei2023channel,dardari2021access} have assumed the array to be spatially continuous, i.e., edge-to-edge antenna deployment with zero antenna spacing or infinitely large numbers of infinitesimal antennas. This  configuration enhances the performance but also causes high fabrication complexity and complicated inter-antenna coupling. By contrast, discrete-aperture XL-MIMO with half-wavelength spacing was studied in \cite{lu2021communicating,lu2022mutiUser,li2022modular,li2022modularLong}. However, these works did not adopt the EM channel model and therefore the impact of polarization mismatch could not be analyzed.

Another crucial attribute of XL-MIMO is the spatial non-stationarity\cite{de2020non}. Due to the large array dimension, different parts of the array might have different views of the propagation environment. Besides, due to the variations of channel amplitudes and incident wave directions across the whole array, the power of the signal transmitted by a user may be received mainly by a portion of the array, which motivates the notion of user visibility region (VR). The existence of spatial non-stationarities was validated by experimental measurements in \cite{payami2012channel}. The authors of \cite{han2020channel} proposed a near-field channel estimation algorithm for XL-MIMO, which  also estimates the mapping between VRs and users. The system performance in the presence of VRs was
analyzed in \cite{li2015capacity,ali2019linear,yang2020uplink,xu2023low}. By exploiting the feature that users at different locations may possess different VRs,  novel algorithms were proposed, in terms of low-complexity detectors \cite{amiri2018extremely}, random access and user scheduling\cite{yang2020uplink,marinello2022exploring,nishimura2020grant}, and antenna selection \cite{marinello2020antenna}. Nevertheless, significant research gaps remain. Firstly, most of the existing contributions assumed that the VR information was available for algorithm design. A rigorous VR detection algorithm has not been reported yet. Secondly, for tractability, the uniform linear array (ULA) model was widely adopted in these works. For the  general  uniform planar array (UPA) model,   VR detection and  the overlapping relationship between VRs of different users are more complex and challenging. Finally, most of the existing works assumed that the antennas outside the VR do not receive any signal. In practice, these antennas  receive small but not zero power, which complicates algorithm design.

To fill the above research gaps, this work investigates discrete-aperture XL-MIMO based on an EM channel model. We first analyze the impact of the near-field channel on the theoretical performance, which sheds light on the effect of spatial non-stationarities and  motivates the design of a VR detection algorithm. Based on the obtained VR information,  two low-complexity symbol detection algorithms for XL-MIMO are proposed.  The main contributions of this paper are listed as follows.

\begin{itemize}
	\item Based on the EM channel model, we derive an explicit expression of the signal-to-noise-ratio (SNR) for discrete-aperture XL-MIMO with a single user. We analytically study the near-field characteristics of the SNR, and provide insights into the impact of the discrete aperture and  polarization mismatch. 
	We also complement the classical Fraunhofer distance to encompass the impact of arbitrary signal incident directions and the power variations across the whole array in the near field.

	\item For multi-user XL-MIMO systems, conventional whole array (WA)-based symbol detectors are proposed and analyzed, including maximum ratio combining (MRC), zero-forcing (ZF), and minimum mean-squared error (MMSE) detection. Next, inspired by the insights drawn from the single-user scenario, we propose a sub-arrays-based VR detection algorithm based on the explicit expressions of the received signal power. Then, a VR-based low-complexity linear symbol detection algorithm is proposed.
	
	\item To further exploit the VR information and to reduce complexity, we propose a graph theory-based user partition algorithm. The users whose VRs overlap exceeds a certain threshold are partitioned into one group. Then, the partial ZF (PZF) detector is utilized to eliminate only the interference within the group. 
	
	\item Simulation results are provided to validate the correctness of analytical results and reveal that the proposed algorithms achieve very similar performance as the conventional WA-based design but with much reduced complexity.
	
\end{itemize} 

The remainder of this paper is organized as follows.  Section \ref{section2} provides the EM channel model for near-field wireless systems. Section \ref{section3} derives the closed-form SNR expression for single-user transmission and analyzes the impact of the discrete aperture and polarization mismatch. Section \ref{section4} proposes the VR detection algorithm, the user partition algorithm, and two low-complexity symbol detection algorithms. Section \ref{section5} presents the numerical results and Section \ref{section6} concludes the paper.

\emph{Notations}: Vectors and matrices are denoted by boldface lower case and upper case letters, respectively. The transpose, conjugate transpose, and inverse of matrix $\bf X$ are denoted by ${\bf X}^T$, ${\bf X}^H$, and ${\bf X}^{-1}$, respectively. $\left[{\bf X}\right]_{(:,k)}$ denotes the $k$-th column of matrix $\bf X$.  $\mathcal{O}$ denotes the standard big-O notation. $|\mathcal{B}|$ denotes the cardinality of set $\mathcal{B}$. $\nabla_{\boldsymbol{x}} \times$ denotes the curl operation with respect to $\boldsymbol{x}$. Besides, $\mathbf{I}_a$ denotes the identity matrix with a dimension of $a\times a$.   The space of $a\times b$ complex matrices is denoted by $\mathbb{C}^{a\times b}$. The $l_2$ norm of a vector $\mathbf{x}$ and the absolute value of a scalar $x$ are denoted by $\left\|\mathbf{x}\right\|$ and $\left|x\right|$, respectively. $j$ denotes the square root of $-1$. $\lfloor a \rfloor$ denotes the nearest integer smaller than $a$.

\section{System Model}\label{section2}
\begin{figure}
	\centering
	\includegraphics[width= 0.45\textwidth]{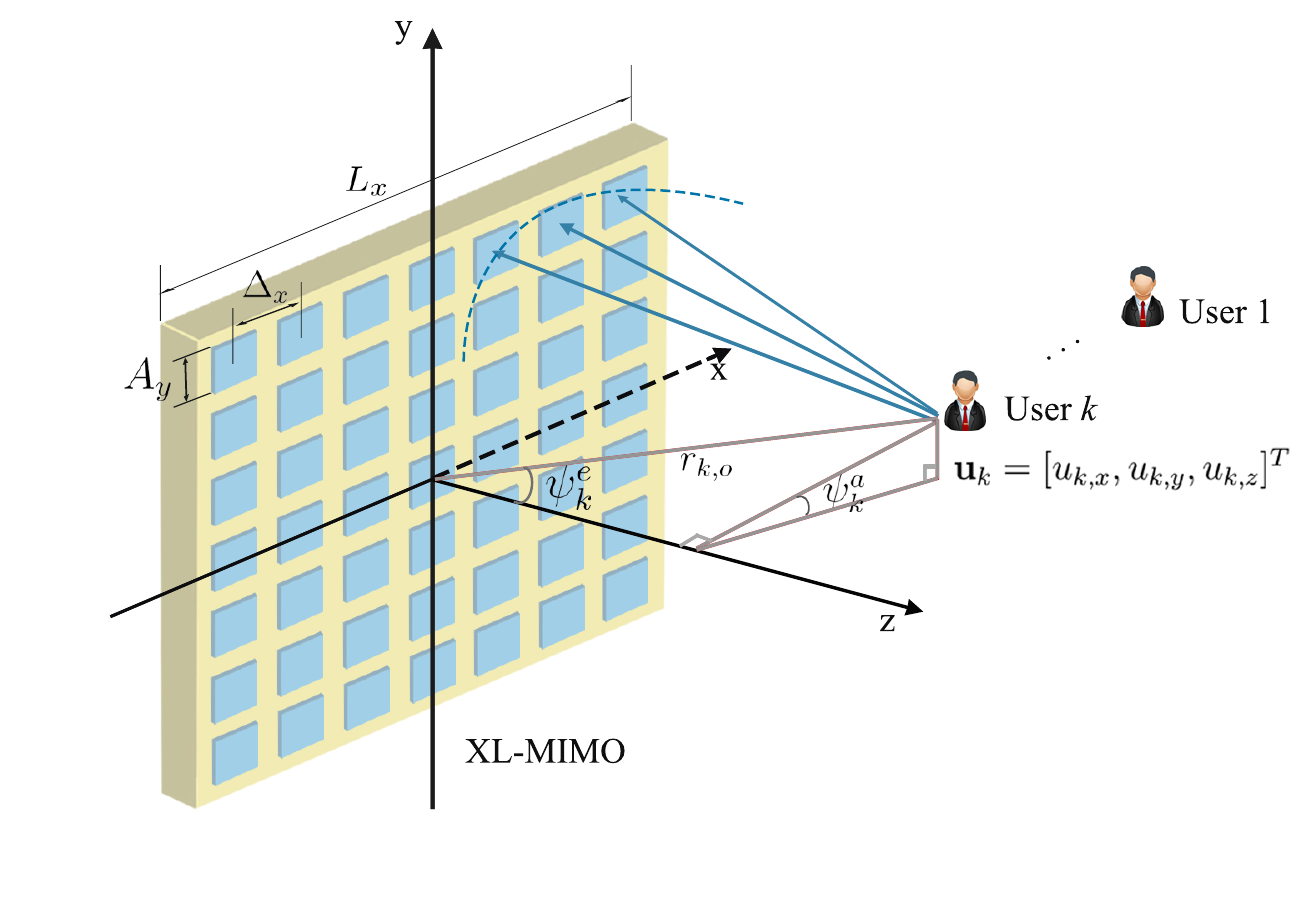}
	\DeclareGraphicsExtensions.
	\caption{Illustration of the considered XL-MIMO system.}
	\label{figure1}
\end{figure}

As illustrated in Fig. \ref{figure1}, we consider the uplink transmission from $K$ single-antenna users to an XL-MIMO array. The user indices are collected in set $\mathcal{K}=\{1,\ldots,K\}$. For tractability, we establish a coordinate system with the center of the array as the origin. The $xoy$ plane is the plane in which the array lies, and the $z$-axis is perpendicular to the array. The location of user $k\in \mathcal{K}$ is denoted by $\mathbf{u}_k=[u_{k,x}, u_{k,y}, u_{k,z}]^T$. 

\subsection{XL-MIMO Array Structure}
Unlike the works considering continuous surfaces or holographic surfaces \cite{Dardari2020LIS,bijoson2020nearField,de2020near,bjornson2021primer,hu2018beyond,pizzo2022holo,wei2023channel}, a discrete-aperture array model is adopted in this work. The XL-MIMO array  consists of  $M=M_x M_y$ antenna elements. The effective area of each antenna element is denoted by $A=A_xA_y$, where $A_x$ and $A_y$ are the lengths of the sides along the $x$- and $y$- axes, respectively\footnote{To facilitate the  analysis, we do not specify the form of the antennas  (wire antennas, aperture antennas, etc.) on the array. We only focus on the effective area of the antenna\cite{balanis2016antenna}, which enables us to establish a unified model. The result  serves as an ideal case which is tractable to characterize the upper-bound performance of the communication system and draw insights.}. The antenna spacing along the $x$- and $y$- axes are denoted by $\Delta_x$ and $\Delta_y$, respectively, with $\Delta_x > A_x$ and $\Delta_y>A_y$. Then, $\eta\triangleq\frac{A}{\Delta_x\Delta_y} \leq 1$ can be referred to as the array occupation ratio characterizing the impact of the discrete aperture of the XL-MIMO array. Considering the $(m_x, m_y)$-th antenna element, the coordinate of its center is given by $\mathbf{p}_{m_x, m_y}=\left[m_x \Delta_x, m_y \Delta_y, 0\right]^T$, $m_c\in \mathcal{M}_c$, where
	\begin{align}\label{range}
		\mathcal{M}_c = \left\{-{(M_c-1)}/{2}, \ldots,{(M_c-1)}/{2}\right\}, c \in\{x, y\}.
	\end{align}
Besides, the region covered by the effective area of the $(m_x, m_y)$-th antenna element is described as $S_{m_x, m_y}=\left[m_x \Delta_x-\frac{{A}_x}{2}, m_x \Delta_x+\frac{ {A}_x}{2}\right] \times\left[m_y \Delta_y-\frac{ {A_y}}{2}, m_y \Delta_y+\frac{{A_y}}{2}\right]$. The distance between user $k$ and the center of the $(m_x, m_y)$-th array element is given by 
\begin{align}\label{rkmxmy}
\begin{aligned}
&{r}_{k,m_x, m_y}=\left\|\mathbf{r}_{k,m_x, m_y}\right\|=\left\|  \mathbf{p}_{m_x, m_x}-\mathbf{u}_k  \right\|\\
&=\sqrt{\left(m_x \Delta_x-u_{k,x}\right)^2+\left(m_y \Delta_y-u_{k,y}\right)^2+\left(u_{k,z}\right)^2}   .
\end{aligned}
\end{align}
 The distance and the  azimuth and elevation angles of arrival (AoAs) from user $k$ to the array center are denoted by $r_{k,o}=\|\mathbf{u}_k\|$  and $\psi_k^a$ and $\psi_k^e$, respectively, where $u_{k,x}=r_{k,o}\sin\psi_k^e\cos\psi_k^a $, $u_{k,y}=r_{k,o}\sin\psi_k^e\sin\psi_k^a $, and $u_{k,z}=r_{k,o}\cos\psi_k^e $.

\subsection{Channel Modeling}
Due to the near-field behavior, it is necessary to distinguish the power and phase of  different elements  when modeling the channel between the user and the XL-MIMO array. Specifically, the channel from  user $k$ to the $(m_x, m_y)$-th antenna element of the XL-MIMO array can be expressed in the following form:
	\begin{align}\label{channel_near_field}
		h_{k,m_x, m_y} = \sqrt{\xi_{k,m_x, m_y}} e^{-j \chi_{k,m_x, m_y} },
\end{align} 
where $ \xi_{k,m_x, m_y} $ and $  \chi_{k,m_x, m_y} $ are the channel power and phase, respectively. By merging $ h_{k,m_x, m_y}$, $\forall m_x ,  m_y$, into a vector, the channel $\mathbf{h}_k\in\mathbb{C}^{M\times 1}$ from user $k$ to the whole array can be obtained.

The Dyadic Green's function-based EM channel model is applied for modeling the power $ \xi_{k,m_x, m_y} $ and the phase $  \chi_{k,m_x, m_y}  $ of the channel $h_{k,m_x, m_y}$ \cite{Poon2005DoF,Dardari2020LIS,bijoson2020nearField}. This channel model is more practical and allows the characterization of the impact of EM polarization effects. Specifically, consider a point $  \mathbf{p}=\left[p_x, p_y, 0\right]^T $  located in the region covered by the $(m_x, m_y)$-th antenna element, i.e.,  $\mathbf{p} \in S_{m_x, m_y}$. Based on Maxwell's equations, the electric field due to the current source of user $k$ satisfies the following inhomogeneous Helmholtz wave equation\cite{Dardari2020LIS}
\begin{align}\label{Helmholtz}
	\left(-\nabla_{\mathbf{u}_k} \times \nabla_{\mathbf{u}_k} \times+k_0^2\right)\boldsymbol{ \mathcal{E}}(\mathbf{u}_k)=j k_0 \kappa \boldsymbol{\mathcal{J}}(\mathbf{u}_k),
\end{align}
where $k_0=\frac{2\pi}{\lambda}$, $\lambda$, and $\kappa$ represent the wavenumber, the wavelength, and the intrinsic impedance of the free space, respectively. $\boldsymbol{\mathcal{E}}(\mathbf{u}_k)$ is the electric field excited by the
current density $ \boldsymbol{\mathcal{J}}(\mathbf{u}_k)$. The inverse mapping of (\ref{Helmholtz}) is given by  $ \boldsymbol{	\mathcal{E}}(\mathbf{p})=\int \boldsymbol{\mathcal{G}}(\mathbf{p}, \mathbf{u}_k) \boldsymbol{{\mathcal{J}}}(\mathbf{u}_k) d \mathbf{u}_k $. In EM theory, $\boldsymbol{	\mathcal{G}}(\mathbf{p},\mathbf{u}_k) $ is referred to as the  Green function which in the radiating near field can be approximately given by\cite{Poon2005DoF}
\begin{align}\label{GreenFunc}
	\boldsymbol{	\mathcal{G}}(\mathbf{p},\mathbf{u}_k) \approx 
	\frac{{-j} \kappa }{2 \lambda\|\mathbf{r}_k\|}\left(\mathbf{I}_3-\hat{\mathbf{r}}_k \hat{\mathbf{r}}_k^H\right)e^{-j \frac{2 \pi}{\lambda}\|\mathbf{r}_k\|},
\end{align}
where $\mathbf{r}_k=\mathbf{p}-\mathbf{u}_k=\left[p_x-u_{k,x}, p_y-u_{k,y},-u_{k,z}\right]^T   $  and $\hat{\mathbf{r}}_k=\frac{\mathbf{r}_k}{\|\mathbf{r}_k\|}$. For notational simplicity, define $r_k = \|\mathbf{r}_k\|$ and hence $ \hat{\mathbf{r}}_k=\left[\frac{p_x-u_{k,x}}{r_k}, \frac{p_y-u_{k,y}}{r_k}, \frac{-u_{k,z}}{r_k}\right]^T $. The Green function $ \boldsymbol{\mathcal{G}}(\mathbf{p},\mathbf{u}_k) $ characterizes the EM response  at   point $\mathbf{p}$ due to the current source at point $\mathbf{u}_k$. The current of the source can be decomposed into different polarization directions as $ \boldsymbol{\mathcal{J}}(\mathbf{u}_k)=\boldsymbol{\mathcal{J}}_x(\mathbf{u}_k) \hat{\mathbf{e}}_x+\boldsymbol{\mathcal{J}}_y(\mathbf{u}_k) \hat{\mathbf{e}}_y+\boldsymbol{\mathcal{J}}_z(\mathbf{u}_k) \hat{\mathbf{e}}_z $, where $\hat{\mathbf{e}}_c$, $c\in\{x,y,z\}$, are the orthonormal basis vectors, and $ \boldsymbol{\mathcal{J}}_c(\mathbf{u}_k) $ denotes the current density in the $c$ polarization direction. As in \cite{Dardari2020LIS,bijoson2020nearField},  low-cost uni-polarized antenna elements are considered in this work and the excited current is assumed to flow in the positive $ y $ direction\footnote{Cross-polarization components are neglected since their intensity could be $30$ dB or more below the primary polarization\cite{balanis2016antenna}.}. Therefore, we have $\boldsymbol{\mathcal{J}}(\mathbf{u}_k)=\boldsymbol{\mathcal{J}}_y(\mathbf{u}_k) \hat{\mathbf{e}}_y$. For unit density $\boldsymbol{\mathcal{J}}_y(\mathbf{u}_k)=1$, we have $\boldsymbol{\mathcal{J}}(\mathbf{u}_k)=[0, 1, 0]^T$ and
\begin{align}\label{greenFun_y}
	\begin{aligned}
		&\boldsymbol{\mathcal{G}}(\mathbf{p}, \mathbf{u}_k)\boldsymbol{ \mathcal{J}}(\mathbf{u}_k) \triangleq \boldsymbol{\mathcal{G}}_y(\mathbf{r}_k) \\
		&=\frac{-{j} \kappa }{2 \lambda r_k}\Bigg[	  \frac{(p_x-u_{k,x})(p_y-u_{k,y})}{r_k^2} ,	1-\left(\frac{p_y-u_{k,y}}{r_k}\right)^2 ,\\
		&\qquad\qquad\frac{(-u_{k,z})(p_y-u_{k,y})}{r_k^2}\Bigg] ^Te^{-j \frac{2 \pi}{\lambda} \|\mathbf{r}_k\|}.
	\end{aligned}
\end{align}
Based on (\ref{greenFun_y}), we can model the channel power between user $k$ and the $(m_x,m_y)$-th antenna element as follows
\begin{align}\label{pathloss1}
	&\xi_{k,m_x, m_y}=\int_{S_{m_x, m_x}} \frac{\lambda^2}{\kappa^2 \pi}\left\|\boldsymbol{\mathcal{G}}_y(\mathbf{r}_k)\right\|^2  \frac{\mathbf{r}_k^T\hat{\mathbf{e}}_z}{\|\mathbf{r}_k\|}d \mathbf{p}\\\label{pathloss2}
	&=\int_{S_{m_x, m_x}} \frac{1}{4 \pi r_k^2} \frac{u_{k,z}}{r_k} \frac{\left(p_x-u_{k,x}\right)^2+u_{k,z}^2}{r_k^2} d \mathbf{p}\\\label{pathloss3}
	&\stackrel{\mathbf{p} \approx \mathbf{p}_{m_x, m_y}}{\approx} \frac{A}{4 \pi} \frac{u_{k,z}\left(\left(m_x \Delta_x-u_{k,x}\right)^2+u_{k,z}^2\right)}{\{\left(m_x \Delta_x \!-\! u_{k,x}\right)^2+\left(m_y \Delta_y \!-\! u_{k,y}\right)^2+u_{k,z}^2\}^{\frac{5}{2}}},
\end{align}
where (\ref{pathloss1}) includes a normalized factor  $  \frac{\lambda^2}{\kappa^2 \pi} $ and a projection factor $ \frac{\mathbf{r}_k^T\hat{\mathbf{e}}_z}{\|\mathbf{r}_k\|} $ that projects the signal from the incident direction onto the normal  direction of the array to characterize the effective projected aperture \cite{Dardari2020LIS,bijoson2020nearField}. In (\ref{pathloss3}), since the size of the antenna element is much smaller than the distance $r_k$, all points $\mathbf{p}$ in the region $ S_{m_x, m_x} $ are approximated by the center point $\mathbf{p}_{m_x, m_y}$. Besides, by applying the approximation, $ \mathbf{p} \approx \mathbf{p}_{m_x, m_y} $, to the phase of the Green function in (\ref{greenFun_y}), the phase of the channel between user $k$ and the $(m_x,m_y)$-th antenna element can be obtained as{\footnote{Based on the reciprocity theorem for antennas and EM theory, the obtained channel model also holds for the downlink communication\cite{balanis2016antenna}. }}
\begin{align}\label{channelphase}
\chi_{k,m_x, m_y} \approx \frac{2 \pi}{\lambda}\|\mathbf{p}_{m_x,m_y} - \mathbf{u}_k\|
=\frac{2 \pi}{\lambda} r_{k,m_x,m_y}.
\end{align} 

By substituting the channel power in (\ref{pathloss3}) and the channel phase in (\ref{channelphase}) into the channel expression (\ref{channel_near_field}), the considered EM channel model is established.

From (\ref{pathloss2}), we can observe that the modeled channel power is comprised of three components, including the free-space pathloss $ \frac{1}{4 \pi r_k^2} $, the  projection coefficient $ \frac{u_{k,z}}{r_k}  $, and the polarization mismatch $ \frac{\left(p_x-u_{k,x}\right)^2+u_{k,z}^2}{r_k^2} $ \cite{bijoson2020nearField}. 
Clearly, if the signal is vertically incident, we have $u_{k,x}=u_{k,y}=0$ and $ \frac{u_{k,z}}{r_k}  =1$; if $u_{k,y}=p_y$, then $ \frac{\left(p_x-u_{k,x}\right)^2+u_{k,z}^2}{r_k^2} =1$ and (\ref{pathloss3}) simplifies to the case without polarization mismatch as in \cite[(2)]{lu2021communicating}. Meanwhile, it can be seen that the expression of the channel phase is more complicated than that used in the far field\cite{zhiTwotimescale2022}.
Thus, the utilized EM channel model is more general and allows us to provide insights into the impact of polarization\footnote{The mutual coupling between different antennas and the return loss are neglected by assuming the ideal decoupling technique and hardware design. The general modeling with antenna coupling is valuable and will be left for our future work.}.

\section{SNR Analysis for Single-User Scenario}\label{section3}
To gain useful insights for system design, in this section, we first focus on a single-user scenario, i.e., $K=1$. We assume that only user $k$ exists. Our objective is to derive an explicit expression for the SNR and analyze its properties in the context of near-field transmission. The signal received at the XL-MIMO array is expressed as 
$ \mathbf{y}_k= \sqrt{p} \mathbf{h}_k x_k + \mathbf{n} $,
where $p$ is the transmit power, $x_k\sim \mathcal{CN}(0,1)$ is the transmit symbol of user $k$, and $\mathbf{n}\sim \mathcal{CN}(\mathbf{0}, \sigma^2 \mathbf{I}_M)$. Using an MRC detector, the SNR is calculated as
\begin{align}\label{snr_sum}
\mathrm{SNR}_k = \frac{p}{\sigma^2} \|\mathbf{h}_k\|^2 = \frac{p}{\sigma^2} \sum_{m_x\in  \mathcal{M}_x}\sum_{m_y\in \mathcal{M}_y} \xi_{k,m_x, m_y},
\end{align}
where $\xi_{k,{m_x},{m_y}}$ is given in (\ref{pathloss3}). In the following, we will present an explicit expression of (\ref{snr_sum}) instead of utilizing the double-sum format.
\begin{theorem}
	Considering channel model (\ref{channel_near_field}) and user $k$ with location $\mathbf{u}_k=[u_{k,x}, u_{k,y}, u_{k,z}]^T$, if $u_{k,z}=0$, we have $\mathrm{SNR}_k=0$; otherwise, the SNR is given by
	\begin{align}\label{rate_UPA}
\begin{aligned}
\mathrm{SNR}_k= &\frac{p}{\sigma^2} \frac{\eta}{6 \pi}\Bigg\{ 
F_k\left(\frac{M_y \Delta_y}{2}-{u}_{k,y}, \frac{M_x \Delta_x}{2}-{u}_{k,x}\right)\\
&+F_k\left(\frac{M_y \Delta_y}{2}-{u}_{k,y}, \frac{M_x \Delta_x}{2}+{u}_{k,x}\right) \\
&+F_k\left(\frac{M_y \Delta_y}{2}+{u}_{k,y}, \frac{M_x \Delta_x}{2}-{u}_{k,x}\right)\\
&+F_k\left(\frac{M_y \Delta_y}{2}+{u}_{k,y}, \frac{M_x \Delta_x}{2}+{u}_{k,x}\right)
\Bigg\},
\end{aligned}
	\end{align}
where $F_k(a, b)=\arctan \left(\frac{a}{{u}_{k,z}} \frac{b}{\sqrt{b^2+a^2+{u}_{k,z}^2}}\right)+\frac{{u}_{k,z}}{2} \frac{a}{a^2+{u}_{k,z}^2} \frac{b}{\sqrt{b^2+a^2+{u}_{k,z}^2}}  $.
\end{theorem}

\itshape {Proof:}  \upshape  See Appendix \ref{App_1}.
\hfill $\blacksquare$

Denote $L_x$ and $L_y$ as the physical dimensions of the XL-MIMO array along the $x$- and $y$-axes, respectively. For large $M$, we have $(L_x -M_x\Delta_x)/L_x\approx 0$ and $(L_y -M_y\Delta_y)/L_y\approx 0$.
Therefore the SNR  in (\ref{rate_UPA}) can be further approximated as
	\begin{align}\label{rate_UPA2}
\begin{aligned} 
	\mathrm{SNR}_k=& \frac{p}{\sigma^2}\frac{\eta}{6 \pi}   \Bigg\{
F_k\left(\frac{L_y }{2}-{u}_{k,y}, \frac{L_x }{2}-{u}_{k,x}\right)\\
&+F_k\left(\frac{L_y }{2}-{u}_{k,y}, \frac{L_x }{2}+{u}_{k,x}\right) \\
&+F_k\left(\frac{L_y }{2}+{u}_{k,y}, \frac{L_x }{2}-{u}_{k,x}\right)\\
&+F_k\left(\frac{L_y }{2}+{u}_{k,y}, \frac{L_x }{2}+{u}_{k,x}\right)
\Bigg\}.
\end{aligned}
\end{align}

The SNR in (\ref{rate_UPA2}) is a function of array lengths $L_x$, $L_y$, user location $\mathbf{u}_k$, and array occupation ratio $\eta=\frac{A}{\Delta_x\Delta_y}\leq 1$. Coefficient $\eta$ highlights the difference of discrete-aperture arrays compared with continuous-aperture arrays. The SNR is an increasing function of $\eta$ since $\eta L_xL_y$ represents the effective array aperture. Next, we aim to analyze the impact of polarization mismatch on the SNR when using a discrete array. Substituting  $ \frac{\left(p_x-u_{k,x}\right)^2+u_{k,z}^2}{r_k^2} =1$ into (\ref{pathloss2}), we can obtain the SNR without polarization mismatch as follows
	\begin{align}\label{rate_wo}
\begin{aligned}
		\mathrm{SNR}_k^{w / o}=&\frac{p}{\sigma^2}\frac{\eta}{6 \pi}\Bigg\{ 
F_k^{w / o}\left(\frac{L_y}{2}-u_{k,y}, \frac{L_x}{2}-u_{k,x}\right)\\
&+F_k^{w / o}\left(\frac{L_y}{2}+u_{k,y}, \frac{L_x}{2}-u_{k,x}\right) \\
&+F_k^{w / o}\left(\frac{L_y}{2}-u_{k,y}, \frac{L_x}{2}+u_{k,x}\right)\\
&+F_k^{w / o}\left(\frac{L_y}{2}+u_{k,y}, \frac{L_x}{2}+u_{k,x}\right)
\Bigg\} ,
\end{aligned}
	\end{align}
 where $F_k^{w / o}(a, b)=\frac{3}{2} \arctan \left(\frac{a b}{u_{k,z} \sqrt{a^2+b^2+u_{k,z}^2}}\right)$. Note that by dividing each term in (\ref{rate_wo}) by $r_{k,o}$ and switching the $ x $-axis and $ y $-axis, this result is identical to that in \cite[(12)]{lu2021communicating}. Comparing (\ref{rate_UPA2}) with (\ref{rate_wo}), we note that function  $F_k(a, b)$ is more complex than $F_k^{w / o}(a, b)$, rendering the theoretical analysis more challenging.
 Recall the assumption of a $ y $-axis polarization direction. As a result, the polarization mismatch escalates with the divergence between the $ y $-coordinate of the user and the antenna element. Thus, when $L_y\to\infty$ (in the severe near field), we can show that $\mathrm{SNR}_k\to\frac{2}{3}\mathrm{SNR}_k^{w / o}$, which  unveils the possible performance degradation attributed to polarization mismatch. Besides, for large $u_{k,z}$  (in the far field), we have $  \arctan (\frac{a}{{u}_{k,z}} \frac{b}{\sqrt{b^2+a^2+{u}_{k,z}^2}}) \approx \frac{ab}{u_{k,z}\sqrt{b^2+a^2+{u}_{k,z}^2}} $ and $ \frac{{u}_{k,z}}{2} \frac{a}{a^2+{u}_{k,z}^2} \frac{b}{\sqrt{b^2+a^2+{u}_{k,z}^2}} \approx  \frac{ab}{2u_{k,z}\sqrt{b^2+a^2+{u}_{k,z}^2}}  $, which results in $F_k(a, b) \approx F_k^{w / o}(a, b)$ and accordingly $\mathrm{SNR}_k \approx \mathrm{SNR}_k^{w / o} \approx \frac{p}{\sigma^2}  \frac{1}{4 \pi r_{k,o}^2} M A \cos \psi_k^e$ \cite{lu2021communicating}. This shows that the polarization mismatch plays a more important role in the near field, and its impact is highly dependent on parameter $a$, i.e., $L_y$ and $u_{k,y}$.

In addition, it is worth noting that the SNR in (\ref{rate_UPA2}) can  be also understood from the perspective of the 3D angles. Define $ \Psi_{k,x}\triangleq\frac{u_{k,x}}{r_{k,o}}=\sin \psi_k^e \cos \psi_k^a $, $\Psi_{k,y}\triangleq \frac{u_{k,y}}{r_{k,o}}=\sin \psi_k^e \sin \psi_k^a$, and $ \Psi_{k,z}\triangleq\frac{u_{k,z}}{r_{k,o}}=\cos \psi_k^e$. Then, the SNR can be reformulated as a function of the AoAs and $r_{k,o}$ as follows:
\begin{align}\label{rate_UPA3}
\begin{aligned} 
	\mathrm{SNR}_k=&\frac{p}{\sigma^2}\frac{\eta}{6 \pi}\Bigg\{
	F_k^A\left(\frac{L_y }{2r_{k,o}}-\Psi_{k,y}, \frac{L_x }{2}-\Psi_{k,x}\right)\\
&	+F_k^A\left(\frac{L_y }{2r_{k,o}}-\Psi_{k,y}, \frac{L_x }{2}+\Psi_{k,x}\right) \\
&	+F_k^A\left(\frac{L_y }{2r_{k,o}}+\Psi_{k,y}, \frac{L_x }{2}-\Psi_{k,x}\right)\\
&	+F_k^A\left(\frac{L_y }{2r_{k,o}}+\Psi_{k,y}, \frac{L_x }{2}+\Psi_{k,x}\right)
\Bigg\},
\end{aligned}
\end{align}
where
$F_k^A(a, b)=\arctan \left(\frac{a}{\Psi_{k,z}} \frac{b}{\sqrt{b^2+a^2+\Psi_{k,z}^2}}\right)+\frac{\Psi_{k,z}}{2} \frac{a}{a^2+\Psi_{k,z}^2} \frac{b}{\sqrt{b^2+a^2+\Psi_{k,z}^2}}  $.

\subsection{Asymptotic Limit}\label{sectionAsymptotic}
In massive MIMO systems operating under far-field conditions and subject to free-space pathloss\cite{bjornson2017massive}, the SNR in the single-user case is given by $	\mathrm{SNR}_k^{\mathrm{far}}=\frac{p}{\sigma^2}  \frac{\lambda^2}{(4\pi)^2r_{k,o}^2} M$, which grows linearly with $M$ towards infinity. However, in reality, the received power cannot exceed the transmitted power based on  energy conservation. In fact, when $M\to\infty$, the near-field channel model has to be applied and the linear scale in $\mathrm{SNR}_k^{\mathrm{far}}$ no longer holds. Based on the EM near-field channel, when $M\to\infty$, the SNR in (\ref{rate_UPA}) converges to
\begin{align}\label{snr_pro}
\mathrm{SNR}_k \rightarrow \frac{p}{\sigma^2}\frac{4 \eta }{6 \pi  } F_k\left(\frac{L_y  }{2}, \frac{L_x  }{2}\right)  \rightarrow \frac{p}{\sigma^2} \frac{\eta}{3} .
\end{align}

If the polarization mismatch is neglected, the asymptotic SNR becomes
\begin{align}\label{snr_noPolari}
	\mathrm{SNR}_k^{w/o} \rightarrow \frac{p}{\sigma^2} \frac{ \eta  }{ \pi} F_k^{w/o}\left(\frac{L_y  }{2}, \frac{L_x  }{2}\right) \rightarrow\frac{p}{\sigma^2} \frac{\eta}{2}  .
\end{align}

If the array is assumed to be continuous i.e., ${A_x}=\Delta_x$ and $ A_y=\Delta_y$, we have
\begin{align}\label{snr_continu}
	\mathrm{SNR}_k^{\mathrm{cont}} \rightarrow  \frac{p}{\sigma^2} \frac{1}{3} .
\end{align}

The results in (\ref{snr_pro}) - (\ref{snr_continu}) provide the practical performance limit for XL-MIMO with an infinitely large array area. 
As can be observed, the SNR in (\ref{snr_pro}) is smaller than the other two cases in (\ref{snr_noPolari}) and (\ref{snr_continu}). In (\ref{snr_noPolari}), at most  half of the power can be received by an infinitely large array surface, since it can capture only the power emitting into  half of the space. With polarization mismatch, the limit reduces from $1/2$ to $1/3$. This is because as $L_y$ increases, the attenuation of the amplitude from the source to the edge of the array becomes severer in the presence of  polarization mismatch, and therefore an additional loss is caused. For the considered discrete aperture, the asymptotic SNR performance is additionally constrained by the array occupation ratio $\eta$ since it characterizes the  effective aperture of the array capable of receiving a signal.

The reason why the SNR is limited as $M\to\infty$ can be explained more clearly from the perspective of geometric views. Consider a user  located perpendicular to the center of the array, i.e., $u_{k,x}=u_{k,y}=0$. Then, the SNR becomes
\begin{align}\label{Fp}
	\begin{array}{l}
		\mathrm{SNR}_k^\mathrm{p}=\frac{p}{\sigma^2}\frac{2 \eta}{3 \pi} \Bigg\{\arctan \left(\frac{L_y/2}{u_{k,z}} \frac{L_x/2}{\sqrt{\left(\frac{L_x}{2}\right)^2+\left(\frac{L_y}{2}\right)^2+u_{k,z}^2}}\right)\\
		+\frac{1}{2} \frac{u_{k,z}}{\sqrt{ \left(\frac{L_y}{2}\right)^2 +u_{k,z}^2}} \frac{L_y/2}{\sqrt{\left(\frac{L_y}{2}\right)^2 +u_{k,z}^2}} \frac{L_x/2}{       \sqrt{\left(\frac{L_x}{2}\right)^2+\left(\frac{L_y}{2}\right)^2+u_{k,z}^2}  }\Bigg\}.
	\end{array}
\end{align}

\begin{figure}[t]
	\centering
	\includegraphics[width= 0.4\textwidth]{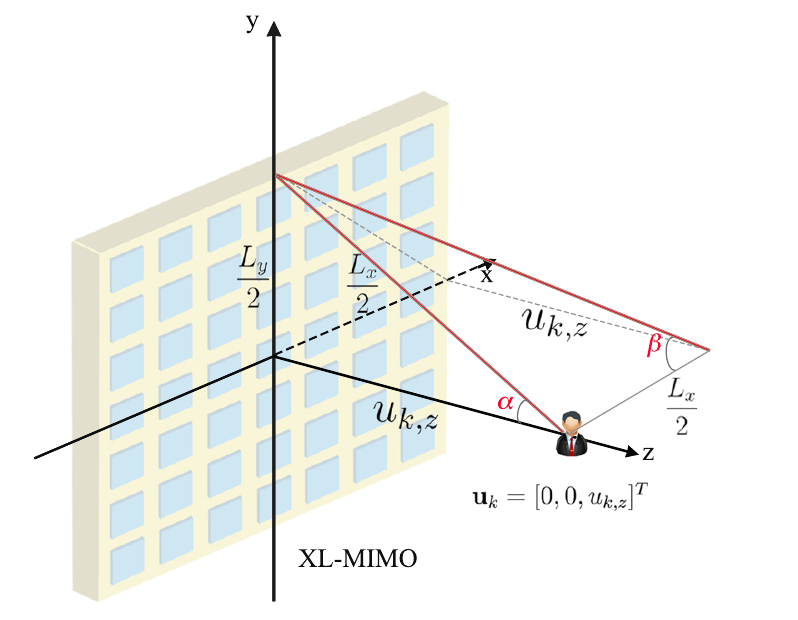}
	\DeclareGraphicsExtensions.
	\caption{Geometric interpretation of the SNR.}
	\label{figure2}
\end{figure}
Eq. (\ref{Fp}) can be reformulated in the form of geometric angles. As illustrated in Fig. \ref{figure2}, we  define two angles $\alpha$ and $\beta$ so that $\tan \alpha = \frac{L_y/2}{u_{k,z}}$ and $\cos \beta =\frac{L_x/2}{\sqrt{\left(\frac{L_x}{2}\right)^2+\left(\frac{L_y}{2}\right)^2+u_{k,z}^2}} $. Then, we have
\begin{align}\label{rate_geo}
\mathrm{SNR}_k^\mathrm{p}=\frac{p}{\sigma^2}\frac{2 \eta}{3 \pi}  \left\{\arctan (\tan \alpha \cos \beta)+\frac{1}{2} \sin {\alpha} \cos \alpha \cos \beta\right\}.
\end{align}
The SNR in (\ref{rate_geo}) is a function of angles $\alpha$ and $\beta$.  Even as the array aperture expands to an infinitely large extent,  the angles of view from the user to the array, i.e., $\alpha$ and $\beta$, remain limited. Specifically, if $L_x \to\infty$, we have $\beta\to0$. When $L_y\to\infty$, we have $\alpha, \beta\to \frac{\pi}{2}$. If both $L_x$ and $L_y$ tend to infinity, the angles converge to $\alpha\to \frac{\pi}{2}$ and $\beta\to\frac{\pi}{4}$. As a result, the SNR is limited by the angles of view and cannot increase unboundedly.

\subsection{XL-ULA}
To shed more light on the impact of polarization mismatch, in this section, we consider a simplified case where the XL-UPA simplifies to an XL-ULA, i.e., $M_x=1$ or $M_y=1$. 

\begin{theorem}
When $M_y=1$, the SNR for the XL-ULA simplifies to
\begin{align}\label{rate_ULA}
\begin{aligned}
\mathrm{SNR}_k^{\mathrm {ULA }}=&\frac{p}{\sigma^2} \frac{A }{4 \pi \Delta_x}\Bigg\{F_k^{\mathrm {ULA }}\left(\frac{M_x \Delta_x}{2}-u_{k,x}\right)\\
&+F_k^{\mathrm {ULA }}\left(\frac{M_x\Delta_x}{2}+u_{k,x}\right)\Bigg\},
\end{aligned}
\end{align}
where $ F_k^{\mathrm {ULA }}(a)=\frac{a\left(a^2 u_{k, y}^2+3 u_{k, z}^2\left(a^2+u_{k, y}^2+u_{k, z}^2\right)\right) u_{k, z}}{3\left(u_{k, y}^2+u_{k, z}^2\right)^2\left(a^2+u_{k, y}^2+u_{k, z}^2\right)^{\frac{3}{2}}} $.

\end{theorem}

\itshape {Proof:}  \upshape  See Appendix \ref{App_2}.
\hfill $\blacksquare$

The result in (\ref{rate_ULA}) is consistent with \cite{lu2021communicating} only if we have $u_{k,y}=0$. This is because when $M_y=1$, the $ y $-coordinate of all antenna elements of the XL-ULA is $0$. When the $ y $-coordinate of the user is also $u_{k,y}=0$, the difference of $ y $-coordinate  of the user and all antennas elements becomes zero, and therefore, there will be no polarization mismatch in pathloss (\ref{pathloss3}). For $M_x\to\infty$, the asymptotic limit of the SNR in (\ref{rate_ULA}) is given by
\begin{align}\label{rate_ULA_utimate}
	\mathrm{SNR}_k^{\mathrm {ULA }} \stackrel{M \rightarrow \infty}{\rightarrow} \frac{p}{\sigma^2} \frac{A}{2 \pi \Delta_x} \frac{u_{k,z}\left(u_{k,y}^2+3 u_{k,z}^2\right)}{3\left(u_{k,y}^2+u_{k,z}^2\right)^2}.
\end{align}
As can be seen,  (\ref{rate_ULA_utimate}) is independent of  $u_{k,x}$. This is because as $M_x\to\infty$, the array attains an infinite length, rendering the user's $x$-coordinate inconsequential.   Furthermore, the asymptotic SNR for the XL-ULA without polarization mismatch is given by
\begin{align}\label{rate_ULA_utimate_without}
\begin{aligned}
	&\mathrm{SNR}_{k,\mathrm{w/o}}^{\mathrm {ULA }} \stackrel{M \rightarrow \infty}{\rightarrow} \frac{p}{\sigma^2}\frac{A}{2 \pi \Delta_x} \frac{u_{k,z}}{u_{k,y}^2+u_{k,z}^2} \\
	&=\frac{p}{\sigma^2}\frac{A}{2 \pi \Delta_x} \frac{u_{k,z}\left(3 u_{k,y}^2+3 u_{k,z}^2\right)}{3\left(u_{k,y}^2+u_{k,z}^2\right)^2}\geq  \mathrm{SNR}_k^{\mathrm {ULA }} ,
\end{aligned}
\end{align}
and the gap between (\ref{rate_ULA_utimate}) and (\ref{rate_ULA_utimate_without}) is
\begin{align}\label{gap_snr}
\begin{aligned}
	&D_{\mathrm{SNR}_k}=  \mathrm{SNR}_{k,\mathrm{w/o}}^{\mathrm {ULA }}  - \mathrm{SNR}_{k}^{\mathrm {ULA }}  =\frac{p}{\sigma^2}\frac{A}{ 3\pi \Delta_x} \frac{u_{k,z} u_{k,y}^2}{\left(u_{k,y}^2+u_{k,z}^2\right)^2},
\end{aligned}
\end{align}
which first increases and then decreases with respect to $u_{k,y}$. Specifically, we have $ D_{\mathrm{SNR}_k} \rightarrow 0$ as $u_{k,y} \rightarrow 0  $ and    $u_{k,y} \rightarrow \infty $. This is because when $u_{k,y}=0$, the user possesses the same $ y $-coordinate as the whole ULA, and therefore the polarization mismatch vanishes. As $u_{k,y}$ increases, the discrepancy in the $ y $-coordinate widens, leading to a larger polarization mismatch. For large enough $u_{k,y}$, the user will be located in the far field and therefore the gap vanishes.

\begin{figure}[t]
	\centering
	\centering
	\includegraphics[width= 0.35\textwidth]{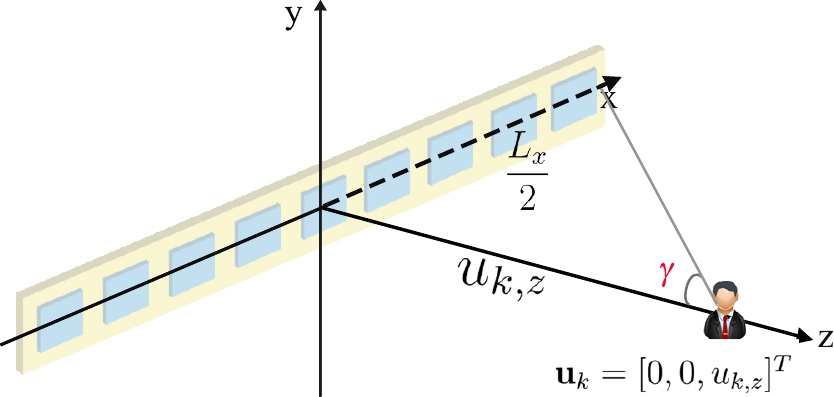}
	\DeclareGraphicsExtensions.
	\caption{Geometric interpretation in the case of XL-ULA.}
	\label{figure3}
\end{figure}
Besides, for $u_{k,x}=u_{k,y}=0$, we can rewrite the SNR expression in (\ref{rate_ULA}) as follows
\begin{align}
	\begin{aligned}
		\mathrm{SNR}_k^{\mathrm {ULA }}&=\frac{p}{\sigma^2}\frac{A  }{2 \pi\Delta_x u_{k,z}}\Bigg(\frac{{L_x/2}}{\sqrt{(\frac{L_x}{2})^2+u_{k,z}^2}}\Bigg)^3 \\
		& =\frac{p}{\sigma^2}\frac{A }{2 \pi \Delta_xu_{k,z}}(\sin \gamma)^3,
	\end{aligned}
\end{align}
where $\gamma$  denotes half of the view angle from the user to the ULA as illustrated in  Fig. \ref{figure3}. Obviously, when $M_x\to\infty$, we have $\gamma\to\frac{\pi}{2}$, and therefore the SNR is limited. 

\subsection{Near-Field/Far-Field Boundary}
In this section, we examine the boundary differentiating the near and far fields  for the considered discrete array with an EM channel model.  A classic result for distinguishing the near and far fields is the Fraunhofer distance\cite{selvan2017fraunhofer,cui2022channel,zhi2022actice}, i.e., $d_{f}=\frac{2D^2}{\lambda}$, where $D$ is the largest array aperture.  It states that if the distance is larger than $d_f$, the maximal phase error of the received signal across the array resulting from approximating the spherical wavefront to the planar wavefront, would be smaller than $\frac{\pi}{8}$ radians. This result has some limits. Firstly, it is obtained based on the condition $\psi_k^e=\psi_k^a=0$, meaning it applies only to signals impinging perpendicularly on the center of the array. Therefore, the impact of the direction of the incident signal is not captured. Secondly, this result only distinguishes the near and far fields according to the signal phase. However, the near and far fields also differ in the feature of the signal power as analyzed in Section \ref{sectionAsymptotic}. Finding the boundary distinguishing the near and far fields based on the variation of the signal power across the array is also of practical importance. Therefore, a more general field boundary is needed by taking into consideration the impact of incident signal directions and the variation of the signal power across the array. 

In the following, we will first determine the field boundary based on the signal phase (with any incident directions) and signal power, respectively, and finally discuss the general result.

\subsubsection{Near-field region characterized by phase error with arbitrary incident directions} To begin with, we complement the classic Fraunhofer distance by further considering arbitrary incident signal directions (i.e., $\forall \psi_k^e$, $\psi_k^a$). To this end, we commence by deriving the maximal phase error across the array as a function of signal directions.
Recall that the phase of the channel  between user $k$ and the $(m_x,m_y)$-th antenna elements is $\chi_{k, m_x, m_y} = \frac{2 \pi}{\lambda} r_{k, m_x, m_y}$ where the distance $r_{k,m_x,m_y}$ in (\ref{rkmxmy}) can be reformulated as (\ref{rkmxmy2}) shown at the bottom of the page,
\begin{figure*}[b]
	\hrulefill
\begin{align}\label{rkmxmy2}
	\begin{aligned}
		r_{k,m_x,m_y} =r_{k, o} \sqrt{1+\frac{-\left(2 m_x \Delta_x r_{k, o} \Psi_{k, x}+2 m_y \Delta_y r_{k, o} \Psi_{k, y}\right)+\left(m_x \Delta_x\right)^2+\left(m_y \Delta_y\right)^2}{r_{k, o}^2}},
	\end{aligned}
\end{align}
\end{figure*}
with $ \Psi_{k,x}=\frac{u_{k,x}}{r_{k,o}}=\sin \psi_k^e \cos \psi_k^a $ and $\Psi_{k,y}= \frac{u_{k,y}}{r_{k,o}}=\sin \psi_k^e \sin \psi_k^a$. Then, based on the second-order Taylor approximation $ (1+x)^{\frac{1}{2}} \approx 1+\frac{1}{2} x-\frac{1}{8} x^2 $ \cite{cui2021near} and neglecting the terms of  $ \mathcal{O}\left({r_{k,o}^{-2}}\right)  $  and $ \mathcal{O}\left({r_{k,o}^{-3}}\right) $, we obtain (\ref{rkappro}), shown at the bottom of the page.
\begin{figure*}[b]
	\hrule
\begin{align}\label{rkappro}
	\begin{aligned}
		r_{k,m_x,m_y} &\approx 
		r_{k, o}-\left(m_x \Delta_x \Psi_{k, x}+m_y \Delta_y \Psi_{k, y}\right)+\frac{\left(m_x \Delta_x\right)^2\left(1-\Psi_{k, x}^2\right)+\left(m_y \Delta_y\right)^2\left(1-\Psi_{k, y}^2\right)-2 \Delta_x \Delta_y \Psi_{k, x} \Psi_{k, y} m_x m_y}{2 r_{k, o}}.
	\end{aligned}
\end{align}
\end{figure*}

Note that based on the first-order Taylor approximation of $r_{k,m_x,m_y} $ in (\ref{rkmxmy2}), the phase of the channel between user $k$ and the $(m_x,m_y)$-th antenna element {\itshape in the far field} is expressed as $\chi_{k, m_x, m_y} ^{\mathrm{far}}  =    \frac{2\pi}{\lambda}   \left( r_{k, o}-\left(m_x \Delta_x \Psi_{k, x}+m_y \Delta_y \Psi_{k, y}   \right) \right)  $. Then, the 3D near-/far-field boundary characterized by phase error is obtained in the following theorem.
\begin{theorem}\label{theorem3}
	The phase error-based 3D near-/far-field boundary is comprised of the coordinates $\tilde{\mathbf{u}}_k= [\tilde{r}_{k,o}\Psi_{k, x},  \tilde{r}_{k,o}\Psi_{k, y} ,\tilde{r}_{k,o}\Psi_{k, z}  ]$, where the values of $\tilde{r}_{k,o}$, for all directions $ \psi_{k }^e\in[0,\frac{\pi}{2}], \psi_{k }^a\in[0,2\pi]$, are given by (\ref{raydis}) shown at the bottom of the next page.
\begin{figure*}[b]
	\hrule
\begin{align}\label{raydis}
	\tilde{r}_{k,o} =  \frac{\left(\frac{M_x-1}{2} \Delta_x\right)^2\left(1-\Psi_{k, x}^2\right)+\left(\frac{M_y-1}{2} \Delta_y\right)^2\left(1-\Psi_{k, y}^2\right)+2 \Delta_x \Delta_y\left|\Psi_{k, x} \Psi_{k, y}\right| \frac{M_x-1}{2} \frac{M_y-1}{2}}{\lambda/8}.
\end{align}
\end{figure*}
\end{theorem}

\itshape {Proof:}  \upshape Based on (\ref{rkappro}), the fact $\Psi_{k, x}^2,\Psi_{k, y}^2<1$,  and the domains of $m_x,m_y$ in (\ref{range}), the maximal phase error across the whole array is derived as (\ref{Dchi}) shown at the bottom of the next page,
\begin{figure*}[b]
	\hrule
\begin{align}\label{Dchi}
	\begin{aligned}
		D_{\chi_{k, m_x, m_y}} &= \max_{m_x, m_y} \left(\chi_{k, m_x, m_y}  - \chi_{k, m_x, m_y} ^{\mathrm{far}}\right)\\
		&\approx \max_{m_x, m_y} 
		\frac{2\pi}{\lambda}   \frac{\left(m_x \Delta_x\right)^2\left(1-\Psi_{k, x}^2\right)+\left(m_y \Delta_y\right)^2\left(1-\Psi_{k, y}^2\right)-2 \Delta_x \Delta_y \Psi_{k, x} \Psi_{k, y} m_x m_y}{2 r_{k, o}}\\
		&= \frac{ \pi}{\lambda} \frac{\left(\frac{M_x-1}{2} \Delta_x\right)^2\left(1-\Psi_{k, x}^2\right)+\left(\frac{M_y-1}{2} \Delta_y\right)^2\left(1-\Psi_{k, y}^2\right)+2 \Delta_x \Delta_y\left|\Psi_{k, x} \Psi_{k, y}\right| \frac{M_x-1}{2} \frac{M_y-1}{2}}{r_{k, o}} ,
	\end{aligned}
\end{align}
\end{figure*}
where the maximal value is achieved by either  $(m_x,m_y) =( \frac{M_x-1}{2}, \frac{M_y-1}{2})$ or $(m_x,m_y) =( -\frac{M_x-1}{2}, \frac{M_y-1}{2})$. It can be seen that $D_{\chi_{k, m_x, m_y}}$ decreases with $r_{k,o}$ given any directions $(\psi_{k }^e,\psi_{k }^a)$. Therefore, $\tilde{r}_{k,o}$ in (\ref{raydis}) is obtained by letting $D_{\chi_{k, m_x, m_y}} = \frac{\pi}{8}$.
\hfill $\blacksquare$

The distance in (\ref{raydis}) degrades to classic Fraunhofer distance under the perpendicular incident direction, i.e., $(\psi_{k }^e,\psi_{k }^a)=(0,0)$. Specifically, applying $\Psi_{k,x}=\Psi_{k,y} =0$, $\frac{M_x-1}{2} \Delta_x\approx L_x$, and $\frac{M_y-1}{2} \Delta_y\approx L_y$ into (\ref{raydis}), we have $ \tilde{r}_{k,o} ^{\mathrm{p}}  \approx \frac{8}{\lambda}\left\{\left(L_x/2\right)^2+\left(L_y/2\right)^2\right\}= 2 D^2/ \lambda$ which is identical to the classical result in \cite{selvan2017fraunhofer}. Theorem \ref{theorem3} proves that for any signal incident directions, the near-field distance $\tilde{r}_{k,o} $ always  increases with $M$, i.e., the physical dimensions of the array. Besides, with the square array of $M_x\Delta_x=M_y \Delta_y \approx L$, we have
\begin{align}\label{raydis_square}
\begin{aligned}
&\tilde{r}_{k,o}^{\mathrm{s}} \approx \frac{2L^2}{\lambda}\left(2-\left(\left|\Psi_{k, x}\right|-\left|\Psi_{k, y}\right|\right)^2\right)\\
&= \frac{D^2}{\lambda}\left(2-\left|\sin\psi_{k }^e\right|^2\left(\left|\cos\psi_{k }^a\right|-\left|\sin\psi_{k }^a\right|\right)^2\right)\leq \frac{2 D^2}{\lambda}.
\end{aligned}
\end{align}

(\ref{raydis_square}) reveals that the near-field distance with different incident directions is smaller than the classical Fraunhofer distance. Besides, it can be found that $ \tilde{r}_{k,o}^{\mathrm{s}} $ decreases with $\psi_{k }^e$. Therefore, given $\psi_{k }^a$, the phase error-based near-field region shrinks as $\psi_{k }^e$ increases from $0$ to $\pi/2$, which showcases the impact of the incident signal direction on the field boundary. 
\subsubsection{Near-field region characterized by power variations} As shown in (\ref{snr_sum}), with the MRC, it is the power variation across the array differentiating the SNR performance in the near field  and far field. Therefore, it is meaningful to identify the near-field region in which the power variation across different antenna elements is non-negligible. Inspired by \cite{lu2021communicating}, the degree of the variation of the channel powers across the whole array based on the EM channel model can be quantified as 
\begin{align}\label{vuk}
	\mathrm{v}(\mathbf{u}_k)=
	\frac{\min _{m_x, m_y}  \xi_{k,m_x, m_y}}{\max _{m_x, m_y} \xi_{k,m_x, m_y}}.
\end{align}

For $\mathbf{u}_k$ located in the far field with planar wavefront, the power variation across the array can be neglected and  we have $ \mathrm{v}(\mathbf{u}_k) \approx 1$. As the user moves closer to the array, the near-field behavior manifests itself, and the variation of the power across the array becomes non-negligible.
Intuitively, the value of $ \mathrm{v}(\mathbf{u}_k)$ decreases from $1$ to $0$ as the user moves closer and closer to the array. Therefore,  we can determine the near-field region by finding the locations of $\mathbf{u}_k$ so that $ \mathrm{v}({\mathbf{u}}_k )<\bar{\mathrm{v}}_t $, where $\bar{\mathrm{v}}_t$ is a threshold explaining the maximal acceptable degree of power difference across the whole array in the far field. 

To avoid the exhausting search across the array, in the following, we will provide the closed-form expressions of the variables $(m_x,m_y)$ in (\ref{vuk}) which  maximize and minimize $\xi_{k,m_x, m_y}$ given $\mathbf{u}_k$, respectively.  Based on (\ref{pathloss3}), we can find that the power $\xi_{k,m_x, m_y}$  between user $k$ and the $(m_x,m_y)$-th antenna element decreases with  their $ y $-coordinate difference $\left|m_y\Delta_y-u_{k,y}\right|$ but it is not monotonic of their $ x $-coordinate difference $\left|m_x\Delta_x-u_{k,x}\right|$. This is because when  $\left|m_y\Delta_y-u_{k,y}\right|$ increases, both the distance and the polarization mismatch increase. By contrast, when $\left|m_x\Delta_x-u_{k,x}\right|$ increases, the distance increases but the relative polarization mismatch decreases. By defining $s=\left(m_x \Delta_x-u_{k,x}\right)^2+u_{k,z}^2$ and $v=\left(m_y \Delta_y-u_{k,y}\right)^2$, we can rewrite $\xi_{k,m_x, m_y}$ in (\ref{pathloss3}) as $f_{\xi}(s)=\frac{s}{(s+v)^{\frac{5}{2}}}$ with $  f_{\xi}^{\prime}(s)=(s+v)^{\frac{-7}{2}} {(v-\frac{3}{2} s)}$. For notational simplicity, define $ f_{\text {int}}(a)=\left\lfloor a+\frac{1}{2}\right\rfloor $ as the function rounding $a$ to the nearest integer. Define further $ f_{\pm x}(a)=\frac{u_{k, x}}{\left|u_{k, x}\right|} a $, and $ f_{|\min|}(a, b) $, where $ f_{|\min |}(a, b) =a$ if $ |a|\leq|b| $ and  $ f_{|\min |}(a, b) =b$ if $ |a|>|b| $.
Then, based on the range in (\ref{range}), we can derive the domain of $s\in [ s_{\min},...,s_{\max}]$, where $ s_{\max }=\left(f_{\pm x}\left(\frac{M_x-1}{2}\right) \Delta_x+u_{k, x}\right)^2+u_{k, z}^2 $ and\footnote{For brevity, we assume that $M$ is odd, and the case of even values can be tackled in a similar manner. }
\begin{align}
	s_{\min }=&\bigg(\left\{f_{|\min |}\left(f_{\text {int }}\left(\frac{u_{k, x}}{\Delta_x}\right), f_{\pm x}\left(\frac{M_x-1}{2}\right)\right)\right\} \Delta_x\nonumber\\
	&-u_{k, x}\bigg)^2+u_{k, z}^2.
\end{align}
By analyzing the properties of  $f_{\xi}(s)$, we obtain the following solutions: to maximize $f_{\xi}(s)$, we have $ \overline{m}_y=f_{|\min |}\{f_{\text {int }}(\frac{u_{k, y}}{\Delta_y}), f_{\pm y}(\frac{M_y-1}{2})\}$, $v^*=\left(\overline{m}_y \Delta_y-u_{k,y}\right)^2$,
and
\begin{align}\label{mx_max}
	\overline{m}_x\!=\! \left\{\begin{array}{l}
		f_{\text {int}}\bigg(f_{|\min|}\bigg\{     \frac{\sqrt{\frac{2}{3} v^*-u_{k, z}^2}+u_{k, x}}{\Delta_x},\\ \;\qquad\frac{-\sqrt{\frac{2}{3} v^*-u_{k, z}^2}+u_{k, x}}{\Delta_x}\bigg\}\bigg),\text { if } s_{\min } \leq \frac{2}{3} v^* \leq s_{\max }; \\
		f_{ |\min|}\left\{f_{\text {int}}\left(\frac{u_{k, x}}{\Delta_x}\right), f_{\pm x}\left(\frac{M_x-1}{2}\right)\right\}, \text { if } \frac{2}{3} v^*<s_{\min }; \\
		-f_{\pm x}\left(\frac{M_x-1}{2}\right), \text { if } \frac{2}{3} v^*>s_{\max };
	\end{array}\right.
\end{align}
to minimize $f_{\xi}(s)$, we have $\underline{m}_y=-f_{\pm y}\left(\frac{M_y-1}{2}\right) $, ${v^*}=\left(\underline{m}_y \Delta_y-u_{k,y}\right)^2$, and
\begin{align}\label{mx_min}
	\underline{m}_x=\!\!\left\{\begin{array}{l}
		f_{|\min |}\left\{f_{\text {int}}\left(\frac{u_{k, x}}{\Delta_x}\right), f_{\pm x}\left(\frac{M_x-1}{2}\right)\right\}, \\
		\qquad\text { if } s_{\min } \leq \frac{2}{3} v^* \leq s_{\max }, f_{\xi}\left(s_{\min }\right) \leq f_{\xi}\left(s_{\max }\right); \\
		-f_{\pm x}\left(\frac{M_x-1}{2}\right) , \\
		\qquad\text { if } s_{\min } \leq \frac{2}{3} v^* \leq s_{\max }, f_{\xi}\left(s_{\min }\right)>f_{\xi}\left(s_{\max }\right) ;\\
		f_{|\min |}\left\{f_{\text {int}}\left(\frac{u_{k, x}}{\Delta_x}\right), f_{\pm x}\left(\frac{M_x-1}{2}\right)\right\}, \text { if } \frac{2}{3} v^*>s_{\max }; \\
		-f_{\pm x}\left(\frac{M_x-1}{2}\right) , \text { if } \frac{2}{3} v^*<s_{\min }.
	\end{array}\right.
\end{align}

Based on (\ref{mx_max}) and (\ref{mx_min}), we can calculate $\mathrm{v}(\mathbf{u}_k)$ given $\mathbf{u}_k$.  To find the $\tilde{\mathbf{u}}_k = [ \tilde{u}_{k,x},\tilde{u}_{k,y},\tilde{u}_{k,z}  ]$ on the near/far-field boundary, we can fix $\tilde{u}_{k,x}$ and $\tilde{u}_{k,y}$ and then  use a one-dimensional search to find  the required value of $\tilde{u}_{k,z}$ which leads to $\mathrm{v}(\tilde{\mathbf{u}}_k) = \bar{\mathrm{v}}_t$.

\subsubsection{The general phase- and power-based near-/far-field boundary} 
In the above discussions, we have introduced the phase-based and power-based field boundaries, respectively. These two field boundaries delineate the near-field regions in which the near-field behaviors are evident in terms of the signal phase and signal power, respectively. Consequently, the union of these two regions describes the near-field region beyond which both the power variations and phase errors are slight and the near-field spherical wavefront can be approximated as the far-field planar wavefront with negligible errors. Conversely, the intersection of these two regions characterizes the near-field region inside which both the power variations and phase errors across the whole array are pronounced.

\section{Multi-User Scenario}\label{section4}
Based on the single-user case, the previous section has shed light on the performance and properties of XL-MIMO in the near field.  Next, this section  focuses on the general multi-user scenarios and proposes low-complexity symbol detectors by leveraging the near-field properties.

\subsection{Whole Array-Based Design}
The signal received by the XL-MIMO array from the $K$ users can be expressed as  $ \mathbf{y}=\sqrt{p} \mathbf{H} \mathbf{x} + \mathbf{n}$, where $\mathbf{H}=[\mathbf{h}_1,\ldots,\mathbf{h}_K]$ and $\mathbf{x}=[x_1,\ldots,x_K]$. Define $\overline{\mathbf{H}}_k = [\mathbf{H}]_{(:,\mathcal{K}\backslash k)} $ as  the matrix obtained by excluding the $k$-th column from $\mathbf{H}$. Then, to detect $x_k$, the linear MRC, ZF, and MMSE detectors for user $k$ are given by \cite{brown2012practical,lu2022mutiUser}
 \begin{align}\label{WA_MRC}
 	& \mathbf{a}_{k, \mathrm{MRC}}^H=\frac{\mathbf{h}_k^H}{\left\|\mathbf{h}_k\right\|^2},\\\label{WA_ZF}
 	& \mathbf{a}_{k, \mathrm{ZF}}^H=\frac{\mathbf{h}_k^H \mathbf{P}_k}{\mathbf{h}_k^H \mathbf{P}_k \mathbf{h}_k}, \mathbf{P}_k =\mathbf{I}_M-\overline{\mathbf{H}}_k\left(\overline{\mathbf{H}}_k^{\mathrm{H}} \overline{\mathbf{H}}_k\right)^{-1} \overline{\mathbf{H}}_k^{\mathrm{H}} ,\\\label{WA_MSE}
 	& \mathbf{a}_{k, \mathrm{MMSE}}^H=\frac{\mathbf{h}_k^H \mathbf{R}_k^{}}{\mathbf{h}_k^H \mathbf{R}_k^{} \mathbf{h}_k},\nonumber\\
 	&\mathbf{R}_k=\mathbf{I}_M-\overline{\mathbf{H}}_k\left(\frac{\sigma^2}{p} \mathbf{I}_{K-1}+\overline{\mathbf{H}}_k^{\mathrm{H}} \overline{\mathbf{H}}_k\right)^{-1} \overline{\mathbf{H}}_k^{\mathrm{H}} .
 \end{align}

Based on the detected symbol $\hat{x}_k = \mathbf{a}_{k, \mathrm{C}}^H \mathbf{y}$, $\mathrm{C}\in\{\mathrm{MRC},\mathrm{ZF},\mathrm{MMSE}\}$, the sum user rate is given by $ R=\sum_{k=1}^K\log(1+\mathrm{SINR}_{k,\mathrm{C}})$,
where the SINR of user $k$ is expressed as
\begin{align}\label{mrc}
	& \operatorname{SINR}_{k, \mathrm{MRC}}=\frac{p\left\|\mathbf{h}_k\right\|^2}{p \sum_{i=1, i \neq k}^K\left|\mathbf{h}_k^H \mathbf{h}_i\right|^2 /\left\|\mathbf{h}_k\right\|^2+\sigma^2} ,\\\label{zf}
	& \operatorname{SINR}_{k, \mathrm{ZF}}=\frac{p}{\sigma^2} \mathbf{h}_k^H \mathbf{P}_k \mathbf{h}_k ,\\\label{mmse}
	& \operatorname{SINR}_{k, \mathrm{MMSE}}=\frac{p}{\sigma^2} \mathbf{h}_k^H \mathbf{R}_k \mathbf{h}_k.
\end{align}

For conventional massive MIMO systems employing a ULA array\cite{zhiTwotimescale2022}, the far-field channel of user $k$ can be expressed as ${\mathbf{h}}_k^{\mathrm{far}} =\xi_k\left[1, e^{-j 2 \pi \frac{\Delta_x}{\lambda} \sin \psi_{k }}, \ldots, e^{-j 2 \pi \frac{\Delta_x}{\lambda}(M-1) \sin \psi_{k }}\right]^T$. As can be seen, unlike the considered near-field channel model (\ref{channel_near_field}), the amplitudes of different entries of ${\mathbf{h}}_k^{\mathrm{far}} $ are uniform and the phases of different entries of ${\mathbf{h}}_k^{\mathrm{far}} $ are linearly scaled. Accordingly, the  multi-user interference term in (\ref{mrc}) can be calculated as\cite{bjornson2017massive,zhiTwotimescale2022,zhi2020power}
\begin{align}\label{favorable}
\frac{\left|(\mathbf{h}_k^{\mathrm{far}})^H \mathbf{h}_i^{\mathrm{far}}\right|^2}{\left\|\mathbf{h}_k^{\mathrm{far}}\right\|^2 } =\frac{\sin^2 \left(\pi \frac{\Delta_x}{\lambda} M\left(\sin \psi_{i }-\sin \psi_{k }\right)\right)}{M \sin^2 \left(\pi \frac{\Delta_x}{\lambda}\left(\sin \psi_{i }-\sin \psi_{k }\right)\right)}.
\end{align}
Clearly, if user $k$ and user $i$ do not have the same angle, the interference will tend to zero as $M\to\infty$. Therefore, MRC detectors can achieve rather good performance. However, this favorable property no longer holds for near-field channels characterized by spherical wavefronts, where the amplitudes of different entries of $\mathbf{h}_k$ are different and the phases are no longer linearly scaled either, as shown in (\ref{channel_near_field}). As a result, we have
\begin{align}\label{nonFavor}
\begin{aligned}
&\mathbf{h}_k^H\mathbf{h}_i^{} = \\
&\sum\limits_{{m_x\in\mathcal{M}_x}}^{} {\sum\limits_{{m_y\in\mathcal{M}_y}}^{} {\sqrt {\xi _{k,{m_x},{m_y}}^H{\xi _{i,{m_x},{m_y}}}} } } {e^{j\frac{{2\pi }}{\lambda }\left( {{r_{k,{m_x},{m_y}}} - {r_{i,{m_x},{m_y}}}} \right)}},
\end{aligned}
\end{align}
which cannot be simplified as a function of the difference of the angles between users $k$ and $i$. Meanwhile, the denominator term $ \left\|\mathbf{h}_k\right\|^2 $ remains  finite for large $M$ and therefore the fraction $ {\left|\mathbf{h}_k^H \mathbf{h}_i^{\mathrm{}}\right|^2}/{\left\|\mathbf{h}_k^{\mathrm{}}\right\|^2 }  $ does not tend to zero even if $M\to\infty$. This implies that the low-complexity MRC detector does not work well in XL-MIMO systems in the near field due to the severe interference. To eliminate the interference, ZF or MMSE detectors are necessary, which significantly increases the computational complexity due to the required matrix inversion. To tackle this challenge, in the following, we propose low-complexity ZF/MMSE schemes by exploiting the near-field spatial non-stationarity.

\subsection{VR-Based Low-Complexity Design}
\begin{figure}[t]
	\centering
	\centering
	\includegraphics[width= 0.3\textwidth]{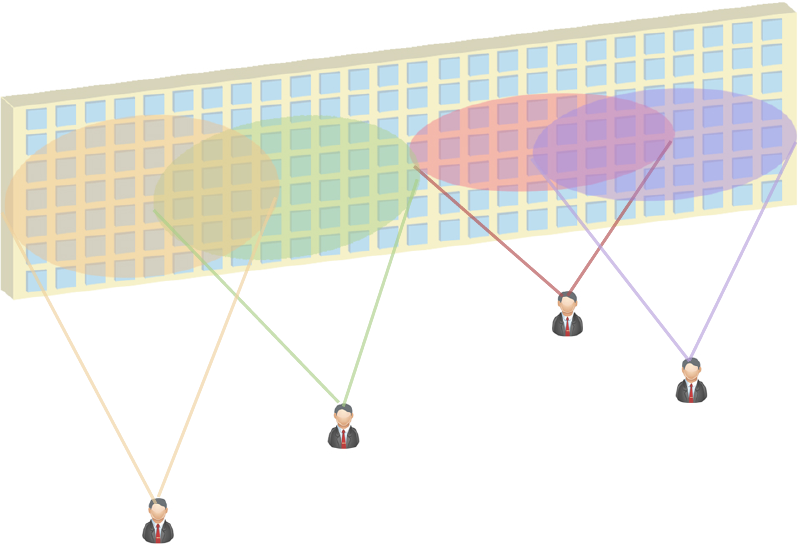}
	\DeclareGraphicsExtensions.
	\caption{Illustration of VRs for different users.}
	\label{figure4}
\end{figure}
Section \ref{section3} has analytically shown that due to the amplitude attenuation across the array, the SNR is limited even for an infinitely large array. In  other words, a limited part of the array receives a large portion of the signal power, which is referred to as the VR of the user, as illustrated in Fig. \ref{figure4}. The VR can be approximated as a combination of sub-arrays of the XL-MIMO array. Considering that the dimensions of the sub-arrays located within the VR of a user could be considerably smaller than the entire array, especially for large $M$, we can utilize the VR of each user to design low-complexity detectors for XL-MIMO.

\begin{lem}\label{lemma1}
	Assume that the XL-MIMO is partitioned into $S=S_x \times S_y$ sub-arrays. For the $(s_x, s_y)$-th sub-array, where $0\leq s_x \leq S_x$ and $0\leq s_y \leq  S_y$, the power of signal received from user $k$ is given by
\begin{align}
\begin{aligned}
	P_{k,s_x,s_y}&=\frac{p}{\sigma^2} \frac{\eta}{6 \pi}\Big\{
F_k\left(f_{s, y, 1} \Delta_y-u_{k, y}, f_{s, x, 1} \Delta_x-u_{k, x}\right)\\
&+F_k\left(-f_{s, y, 1} \Delta_y+u_{k, y}, f_{s, x, 2} \Delta_x-u_{k, x}\right) \\
&+F_k\left(-f_{s, y, 2} \Delta_y+u_{k, y}, f_{s, x, 1} \Delta_x-u_{k, x}\right)\\
&+F_k\left(f_{s, y, 2} \Delta_y-u_{k, y}, f_{s, x, 2} \Delta_x-u_{k, x}\right)
\Big\},
\end{aligned}
\end{align}
		where $f_{s, c, 1}=\frac{-M_c}{2}+s_c \frac{M_c}{S_c}, f_{s, c, 2}=\frac{-M_c}{2}+\left(s_c+1\right) \frac{M_c}{S_c}, c \in\{x, y\}$.
\end{lem}

\itshape {Proof:}  \upshape  For the $(s_x, s_y)$-th subarray, we can obtain the indices of the antennas as  $m_c \in\left\{\frac{-\left(M_c-1\right)}{2}+s_c \frac{M_c}{S_c}, \ldots, \frac{-\left(M_c-1\right)}{2}+\left(s_c+1\right) \frac{M_c}{S_c}-1\right\}$, $c\in\{x,y\}$. The proof follows by deriving the sum of the powers across this sub-array using a similar method as in Appendix \ref{App_1}.
\hfill $\blacksquare$
  \begin{algorithm}[t]
	\caption{ {VR Detection}}
  	\setstretch{1}
	\begin{algorithmic}[1]\label{algorithm1}
		\REQUIRE $\varpi$, $\hat{P}_k=0$, $\mathcal{B}_k=\emptyset$, $\mathrm{SNR}_k$, $ \mathcal{P}_k=\{P_{k,s_x,s_y},\forall s_x, s_y\} $
		\ENSURE $\mathcal{B}_k$
		\FOR{$k=1:K$}
		\STATE Sort $ \mathcal{P}_k$ using quicksort algorithms; $i=1$;
		\WHILE{$\hat{P}_k \le \varpi \mathrm{SNR}_k$}
		\STATE Select the $i$-th element in $ \mathcal{P}_k $ referred to as $ P_{k,s_x^*,s_y^*} $;
		\STATE  $\mathcal{B}_k= \{  \mathcal{B}_k  \cup (s_y^*S_x+ s_x^*) \}$; 
		\STATE $\hat{P}_k=\hat{P}_k+P_{k,s_x^*,s_y^*}$; $i=i+1$;
		\ENDWHILE
		\ENDFOR
	\end{algorithmic}
\end{algorithm}

Based on Lemma \ref{lemma1}, the VR of user $k$ can be determined  by selecting the sub-arrays contributing to the main portion of the received SNR as $ \varpi \mathrm{SNR}_k $, where $\varpi \in [0,1]$ and $\mathrm{SNR}_k$ is given in (\ref{rate_UPA}). The procedure for detecting VR is outlined in Algorithm \ref{algorithm1}, where step  5 collects the sub-array indices for user $k$ in set $\mathcal{B}_k$. Then, we can use only the sub-arrays belonging to $\mathcal{B}_k$ to detect the symbol of user $k$, $\forall k$, which helps reduce the computational complexity. Specifically, for user $k$, we first construct the channel ${\mathbf{H}}_{\mathcal{B}_k} \in \mathcal{C}^{ (\frac{|\mathcal{B}_k| }{S}M)\times K}$ from the $K$ users to the antenna elements belonging to $\mathcal{B}_k$. Then, the symbol of user $k$ can be detected based on the linear detectors in (\ref{mrc})-(\ref{mmse}) by substituting the WA channel matrix $\mathbf{H}$ with VR channel matrix ${\mathbf{H}}_{\mathcal{B}_k} $. Taking VR-based ZF as an example, we have $\mathbf{h}_k^{\mathrm{VR}}=[{\mathbf{H}}_{ \mathcal{B}_k}]_{(:,k)}$, $\overline{\mathbf{H}}_k^{\mathrm{VR}}=[{\mathbf{H}}_{\mathcal{B}_k}]_{(:,\mathcal{K}\backslash k)}$, and the detector of user $k$ is obtained as
\begin{align}\label{VR_ZF}
(\mathbf{a}_{k, \mathrm{ZF}}^{\mathrm{VR}} )^H=\frac{ (   \mathbf{h}_k^{\mathrm{VR}}    )^H \mathbf{P}_k ^{\mathrm{VR}}  }{ ( \mathbf{h}_k^{\mathrm{VR}})^H \mathbf{P}_k^{\mathrm{VR}}   \mathbf{h}_k^{\mathrm{VR}}  },
\end{align}
where
$ \mathbf{P}_k^{\mathrm{VR}}  =\mathbf{I}_{\frac{|\mathcal{B}_k| }{S}M}-\overline{\mathbf{H}}_k^{\mathrm{VR}}  \left(   (\overline{\mathbf{H}}_k^{\mathrm{VR}}) ^{{H}} \overline{\mathbf{H}}_k^{\mathrm{VR}}   \right)^{-1} (\overline{\mathbf{H}}_k^{\mathrm{VR}}) ^{{H}}$.

\subsection{User Partition-Based PZF}
Next, we fully exploit the properties of the VRs to further reduce the computational complexity. In the near field,  users located at different positions may have different VRs, and users whose VRs  do not significantly overlap may  suffer from low mutual interference. Therefore, we can partition the $K$ users into several groups, where the VRs of users in different groups have low overlap.  Then, the users will be mainly affected by the interference caused by the users in their own group. The interference from users in other groups is expected to be weak and can be neglected during detector design. As a result, we propose to utilize the PZF detector \cite{interdonato2020local,zhi2019cache}, which eliminates only intra-group interference and therefore effectively reduces the computational complexity of the required matrix inversion. 
To begin with, we exploit the VR information to propose a user partition algorithm grounded in graph theory.

\begin{definition}\label{lemma2}
\cite{chang2017computing}: A undirected graph can be denoted as $G=(V,E)$, where $V$ and $E$ are the sets of vertices and edges, respectively. $(u,v)\in E$ means that there is an edge between vertices $u$ and $v$. The neighborhood of $u$ is $N(u)=\{v\in V| (u,v)\in E\}$ and the degree of $u$ is $d(u)=|N(u)|$. A  path $ D $ of $ G $ is a degree-two path if all vertices of $ D $  have edges with each other and have degree two. The maximum independent set of graph $G$ is the maximum vertex set $\mathcal{I} \subseteq V$, in which all vertices have no edge. 
\end{definition}

Following Definition \ref{lemma2}, we  construct an undirected graph $G=(V,E)$ with $V=\{v_1,\ldots,v_K\}$ corresponding to the $K$ users. The edge $(v_k,v_i)$ signifies the VR overlap situation between users $k$ and $i$. Specifically, define $ \hat{s}_\mathrm{ovp} \in [0,1] $ as a threshold specifying the maximum acceptable overlap ratio between the VRs of two users. If $\frac{| \mathcal{B}_k  \cap  \mathcal{B}_i | }{ \min\{ |\mathcal{B}_k | , | \mathcal{B}_i | \}} \ge  \hat{s}_\mathrm{ovp}$, the VR overlap ratio between users $k$ and $i$ exceeds the threshold, and we establish an edge $(v_k,v_i)\in E$.  In order to reduce the matrix inversion complexity while guaranteeing the performance, our target is to partition users into as many as groups possible, where the VRs of different groups have low overlap. To this end, based on the graph $G=(V,E)$, we address the following maximum independent set problem:
\begin{align}\label{p1}
		\max \; |\mathcal{I}|,\text { s.t.  }\; \mathcal{I}\subseteq V ;(u, v) \notin E, \forall u, v \in \mathcal{I}.
\end{align}

Upon solving the maximum independent set problem in (\ref{p1}), we obtain as many as possible vertices in $G$ without edges between each other. In other words, we find as many as possible users with low-overlap VRs. Then, each vertex  in $\mathcal{I}$ and its neighborhood constitute a user group.  Based on the user partitioning results, we design the PZF detector to   eliminate only the intra-group interference. The detailed procedure is outlined in Algorithm \ref{algorithm2}. Specifically, steps 2 and 3 construct the graph $G$ based on the VRs. Steps 4-14 find the maximum independent set $\mathcal{I}$ of $G$ exploiting a degree-based reduction algorithm with linear complexity $\mathcal{O}(K)$ \cite{chang2017computing}. Then, each vertex in $\mathcal{I}$ with its neighborhood vertices form one user group $\mathcal{L}$. For each user group, steps 15-19 design the PZF detector by eliminating the interference within the group.

  \begin{algorithm}[t]
	\caption{ User Partition-Based PZF}
	\begin{algorithmic}[1]\label{algorithm2}
		\STATE Initialize $\mathcal{I}=\emptyset$
		\STATE Construct graph $G=(V,E)$, $V=\{v_1,\ldots,v_K\}$, $(v_k,v_i)\in E$  if $| \mathcal{B}_k  \cap  \mathcal{B}_i | \ge \hat{s}_\mathrm{ovp} \min\{ |\mathcal{B}_k | , | \mathcal{B}_i | \}  $
		\STATE Calculate degree $d(v_k)$, $\forall k$; $\mathcal{V}_1 = \{v_k|d(v_k)=1,\forall k\}$, $\mathcal{V}_2 = \{v_k|d(v_k)=2,\forall k\}$, $\mathcal{V}_{3} = \{v_k|d(v_k)>2,\forall k\}$
		\WHILE{$\mathcal{V}_1 $ or $\mathcal{V}_2 $ or $\mathcal{V}_3$  $\neq \emptyset$}
		\IF {$\mathcal{V}_1 \neq \emptyset$ }  
		\STATE Delete the neighborhood of $v_k \in \mathcal{V}_1$ from $G$
		\ELSIF{$\mathcal{V}_2 \neq \emptyset$}
		\STATE Select a vertex $v_k \in \mathcal{V}_2$, find its maximal degree-two path, and delete vertices based on rules in \cite[Lemma 4.1]{chang2017computing}
		\ELSE
		\STATE Delete the vertex with the largest degree
		\ENDIF
		\STATE Update the degree for all vertices 
		\ENDWHILE
		\STATE $\mathcal{I} = \{v_k|d(v_k)=0,\forall k\}$
		\WHILE {  $\mathcal{I}  \neq \emptyset$}
		\STATE  Select $u \in \mathcal{I}$; construct $\mathcal{L} =\{i |v_i\in \{u,N(u)\} \cap i\in \mathcal{K} \}$
		\STATE For users $i$, $i \in \mathcal{L}$, design $ \mathbf{a}_{i, \mathrm{PZF}}^H=\frac{\widehat{\mathbf{h}}_i^H \mathbf{L}_i}{\widehat{\mathbf{h}}_i^H \mathbf{L}_i \widehat{\mathbf{h}}_i}$, 
		$\mathbf{L}_i =\mathbf{I}_{\left|\mathcal{B}_i\right| \frac{ M}{S}}-\widehat{\mathbf{H}}_{\mathcal{L} \backslash i}\left(\widehat{\mathbf{H}}_ {    \mathcal{L} \backslash i
		 }^{\mathrm{H}} \widehat{\mathbf{H}}_{ \mathcal{L} \backslash i   }\right)^{-1} \widehat{\mathbf{H}}_{\mathcal{L} \backslash i}^{\mathrm{H}}  $,
	    where $ \widehat{\mathbf{H}}_{\mathcal{L} \backslash i} = \mathbf{H}(\mathcal{M}_i^{\mathrm{VR}},  \mathcal{L} \backslash i)$, $\widehat{\mathbf{h}}_i = \mathbf{H}(\mathcal{M}_i^{\mathrm{VR}},  i)$, and $\mathcal{M}_i^{\mathrm{VR}}$ denotes the antennas indices in the VR of user $i$
		\STATE Remove $u$ from $\mathcal{I}$
		\ENDWHILE
	\end{algorithmic}
\end{algorithm}

\subsection{Complexity Analysis}
For brevity, only  the complexity of ZF is analyzed since MMSE and ZF have the same asymptotic complexity. We refer to the three algorithms as whole array-based ZF (WA\_LD), VR-based ZF (VR\_LD), and user partitioning-based  PZF (UP\_PZF). The results are presented in Table \ref{tab1}. 
For tractability, we consider an ideal scenario where the VR of each user includes the same number of sub-arrays, i.e., $|\mathcal{B}_k|= |\widetilde{\mathcal{B}|}$, $\forall k$. Then, the number of antennas in the VR of each user reduces from $M$ to $\frac{ |\widetilde{\mathcal{B}|} }{S} M$. We also ideally assume that for the user partitioning algorithm, $K$ users are divided uniformly into $ |\widetilde{\mathcal{I}}| $ groups. Therefore, the  matrix dimension for the inversion operation in the PZF  scheme diminishes from $K$ to $K/{   |\widetilde{\mathcal{I}|}    }$. Table \ref{tab1} reveals that the proposed algorithms can significantly reduce the complexity when $|\widetilde{\mathcal{B}|}/S$ is small and when $|\widetilde{\mathcal{I}|}$ is large.

\begin{table}[t]
	\centering 
	\caption{Computational Complexity}	
	\begin{tabular}{|c|c|c|c|c|c|c}
		\hline
		WA\_ZF&$\mathcal{O}\left\{M^2K^2+K^4\right\}$\\
		\hline
		VR\_ZF&  $\mathcal{O}\{({|\widetilde{\mathcal{B}}| }/{S})^2 M^2K ^2  +K^4  + KS\log(S)\} $\\
		\hline
		UP\_PZF & $\!\!\mathcal{O}\{({|\widetilde{\mathcal{B}}| }/{S})^2M^2{K^2 }/{|\widetilde{\mathcal{I}}|}  \!+\! {K^4}/{|\widetilde{\mathcal{I}}|^3}  \!+\! KS\log(S)\} \!\!$\\
		\hline
	\end{tabular}\label{tab1}
\end{table}

\section{Numerical Results}\label{section5}
In this section, we provide numerical results for validating our analytical conclusions and providing insight into the performance of XL-MIMO systems. Consistent with existing literature \cite{lu2021communicating,lu2022mutiUser,li2022modular,li2022modularLong}, we set $\Delta_x=\Delta_y=\frac{\lambda}{2}=0.0628$ m, $\frac{p}{\sigma^2}= 90$ dB, and  $A=\frac{\lambda^2}{4 \pi}$.

\subsection{Single-User Case}


\begin{figure}[t]
	\setlength{\abovecaptionskip}{0pt}
	\centering
	\includegraphics[width=3.4in]{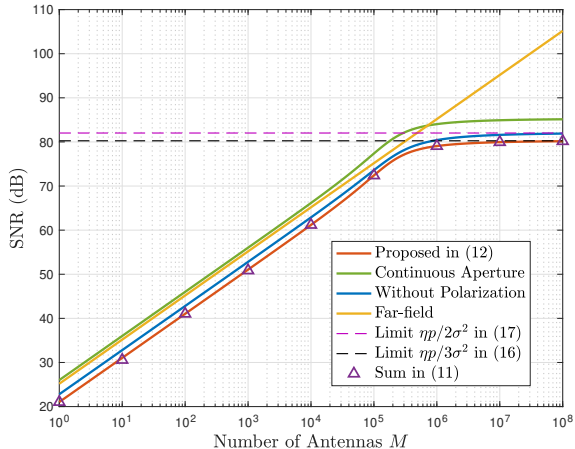}
	\caption{ SNR versus $M$ for XL-UPA, $\mathbf{u}_k=[10 ,10 ,10]$, $M_x=M_y=\sqrt{M}$. }
	\label{figure_1}
\end{figure}
To begin with, the single-user case is considered. Fig. \ref{figure_1} depicts the SNR performance as the aperture of the XL-UPA grows infinitely large.  It can be seen that unlike the far field-based result which increases linearly with $M$, the near field-based SNR initially increases but ultimately saturates as $M$ approaches infinity. Furthermore, the proposed model, which takes into consideration both the discrete aperture and polarization mismatch, characterizes the actual performance with additional loss. Besides, it can be seen that the SNR in (\ref{snr_sum}) which includes a double sum is well approximated by the derived explicit result.

\begin{figure}[t]
	\setlength{\abovecaptionskip}{0pt}
	\centering
	\includegraphics[width=3.4in]{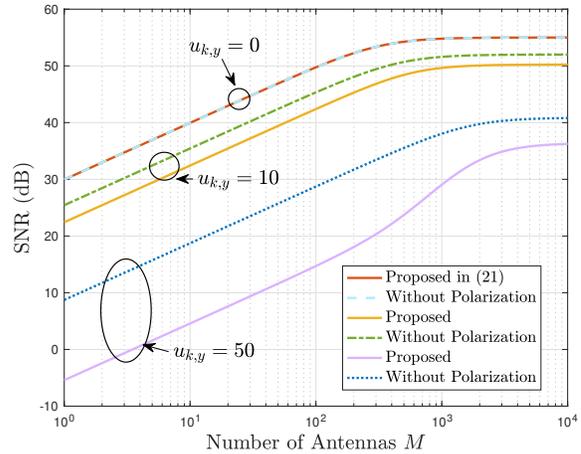}
	\caption{SNR versus $M$ for XL-ULA, $\mathbf{u}_k=[0 ,u_{k,y} ,10]$, $M_x=M$, $M_y=1$.}
	\label{figure_2}
\end{figure}
In Fig. \ref{figure_2}, the asymptotic SNR of XL-ULA is studied. A similar tendency as in Fig. \ref{figure_1} is observed. However, the number of antennas needed for the SNR growth rate to slow down is notably smaller than that in Fig. \ref{figure_1}. The SNR of ULA attains saturation with roughly $10^3$ antenna elements, whereas the required number of antenna elements for UPA is $10^6$.  This is because given a value of $M$, the ULA has a considerably larger dimension than the UPA. Consequently,  the variations in amplitudes,  angles between the signal incident direction and the array normal, and polarization mismatch become more pronounced across the whole array.  As a result, the near-field behavior becomes more obvious for the ULA with a given $M$. This phenomenon can also be understood via the geometric figures in Figs. \ref{figure2} and \ref{figure3}. As $M$ increases,  the enlarging of view of angles $\gamma$ is easier to saturate for the ULA than angles $\alpha$ and $\beta$ for the UPA. Furthermore, it can be observed that the SNR gap between the proposed EM model and the model without polarization mismatch enlarges as the $y$-coordinate of the user increases. This phenomenon agrees with our analytical result (\ref{gap_snr}) since the polarization mismatch is proportional to the difference in the $y$-coordinate between the user and the received antenna. As a result, as $u_{k,y}$ increases, the performance loss caused by polarization mismatch increases, which enlarges the gap.

%

 \begin{figure}[t]
 	\setlength{\abovecaptionskip}{0pt}
 	\centering
 	\includegraphics[width=3.4in]{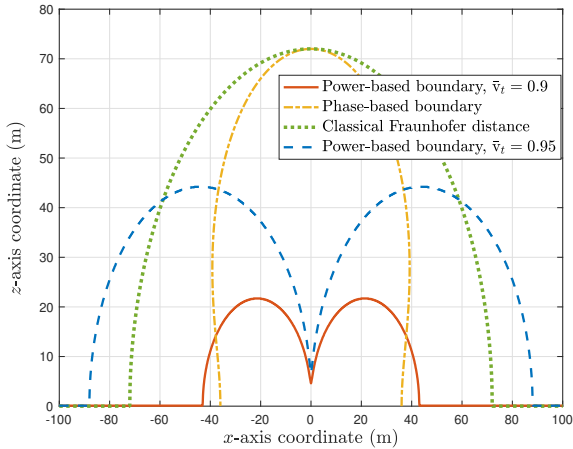}
 	\caption{Field boundaries at $u_{k,y}=0$, $M_x=M_y=25$.}
 	\label{figure_4}
 \end{figure}
  Fig. \ref{figure_4} illustrates the proposed near-/far-field boundaries on the plane of $u_{k,y}=0$ ($\psi_{k }^a=0$ or $\pi$). The threshold $\bar{\mathrm{v}}_t$ is chosen to be $0.9$ or $0.95$ so that the power difference across the whole array can be neglected in the far field. It can be seen that the phase-based boundary considering different signal incident directions is smaller than the classical Fraunhofer distance, and it keeps shrinking as  $\psi_{k }^e$ increases from $0$ to $\pi/2$, which is consistent with our analysis below (\ref{raydis_square}).
Besides, it can be seen that the signal power-based boundary has a different shape from the phase-based boundary.  It shrinks as the user moves towards the center of the array (i.e., as $x$-coordinate $\to 0$). This is because the variation of the power from the center to the  edge of the array is smaller than that from one edge to the other edge of the array. For a milder amplitude variation, the near-field region is reduced. Furthermore, the signal power-based boundary expands with $\bar{\mathrm{v}}_t$ due to the stricter requirement of the degree of the power variation across the array for the far field. Combining the power and phase-based boundaries can provide a useful reference for the practical algorithm design of the XL-MIMO.

\subsection{Multi-User Case}
The previous subsection illustrated the impact of near-field spatial non-stationarities, which inspired the proposed low-complexity design for multi-user scenarios. In this section, numerical results are presented to illustrate the effectiveness of the proposed algorithms. Unless stated otherwise, we assume that $M=10^4$, $M_y=10$, $S_y=2$, $M_x/S_x=10$, $\varpi = 0.8$, and $\hat{s}_{\mathrm{ovp}}=0.6$. $K=20$ users are randomly distributed within the region of $[-25,25]\times [2,12]$ on the $ xoz $ plane.
%

\begin{figure}[t]
	\setlength{\abovecaptionskip}{0pt}
	\centering
	\includegraphics[width=3.4in]{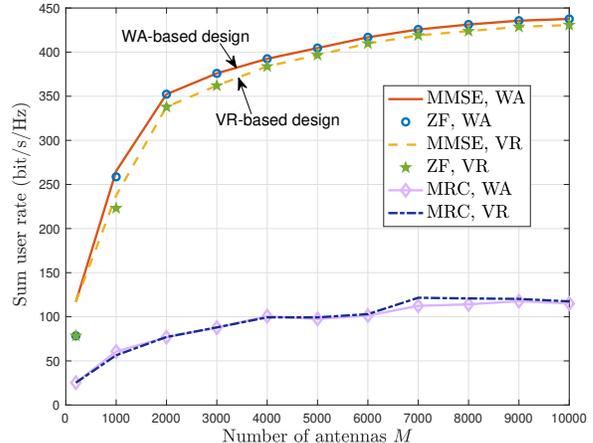}
	\caption{Comparison of WA and VR-based designs.}
	\label{figure_multi_1}
\end{figure}
In Fig. \ref{figure_multi_1}, we  compare the performance of the whole array (WA)-based design with the proposed VR-based design. The WA-based MMSE, ZF, and MRC detection are carried out based on (\ref{WA_MRC})-(\ref{WA_MSE}), while the VR-based design is conducted as (\ref{VR_ZF}). It can be seen that for both the WA and VR-based cases, the sum user rates for the MMSE and ZF detectors coincide for large $M$ and are much higher than that for MRC. This is because in near-field scenarios with spherical wavefronts, the favorable interference condition in (\ref{favorable}) no longer holds, which deteriorates to (\ref{nonFavor}) as a function of distances and angles. Therefore, considering that the computational complexity of ZF and MMSE is much higher than that of MRC, it is necessary to employ low-complexity detectors in XL-MIMO systems.  From Fig. \ref{figure_multi_1}, it can be observed that the proposed VR-based low-complexity MMSE and ZF detectors perform very close to the WA-based design, especially for large $M$. This is because for XL-MIMO with large physical dimensions, the variations in amplitudes,  angles between the signal incident directions and array normal, and polarization mismatch are pronounced across the whole array. As a result, the majority of the signal power is received on a limited portion of the array. In other words, the user will  ``see" only a part of the array (i.e., the VR). If the VRs are accurately detected, it is expected that the proposed VR-based algorithms can achieve performance comparable to that of the WA-based design. Meanwhile, for large $M$, the portion of the array that contributes marginal received power grows. Thus, for a given ratio $\varpi$, the VR detection algorithm (Algorithm \ref{algorithm1}) is more efficient in finding the most relevant sub-arrays, thus diminishing the performance gap between the VR and WA-based designs.

\begin{figure}[t]
	\setlength{\abovecaptionskip}{0pt}
	\centering
	\includegraphics[width=3.4in]{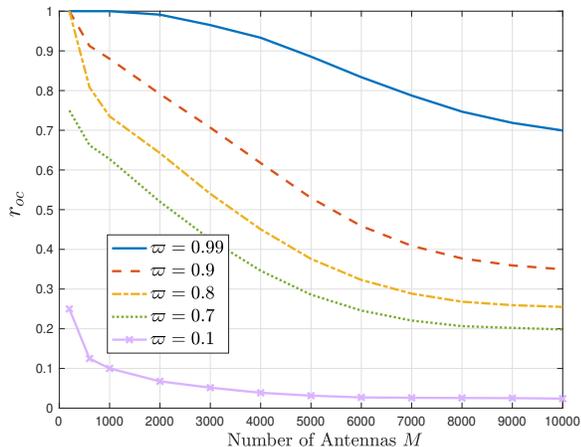}
	\caption{Ratio of average number of antennas employed \\ by VR-based design.}
	\label{figure_multi_2}
\end{figure}

\begin{figure}[t]
	\setlength{\abovecaptionskip}{0pt}
	\centering
	\includegraphics[width=3.4in]{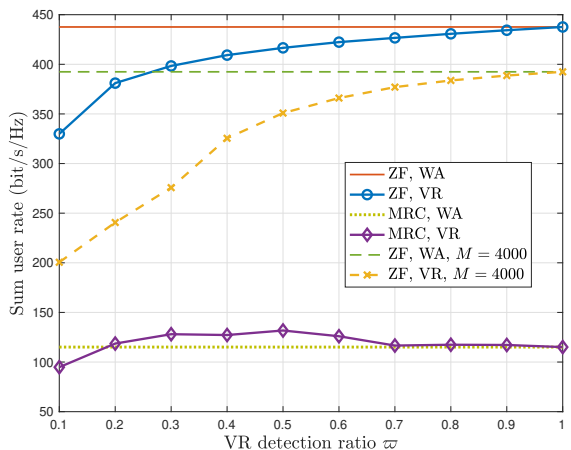}
	\caption{Sum user rate under different  $\varpi$, where $M=10^4$.}
	\label{figure_multi_3}
\end{figure}

Intuitively, the proposed VR-based design aims to achieve a trade-off between performance and complexity,  which can be adjusted by the VR detection ratio $\varpi$ as shown in Algorithm \ref{algorithm1}.  To quantify the complexity, the ratio of the average number of  antennas employed by the VR-based design to the number of antennas employed by the WA-based design is defined as $ {r}_{oc}= \frac{1}{K} \sum_{k\in\mathcal{K}} \frac{|\mathcal{B}_k|}{S}$. Clearly, we have $r_{oc}=1$ for the WA-based design. Next, Figs. \ref{figure_multi_2} and \ref{figure_multi_3} illustrate the trade-off between performance and complexity when utilizing VR-based detectors.

 In Fig. \ref{figure_multi_2}, we observe that ratio $r_{oc}$ is an increasing function of the VR detection ratio $\varpi$, which implies that the number of antennas considered for the computation of VR-based detectors increases with $\varpi $. This is because the VR is defined as the set of sub-arrays that contribute a fraction of $\varpi$ to the total received power. The larger $\varpi$ is, the more sub-arrays are included in the VR. Besides, as can be seen,  $r_{oc}$ is a decreasing function of $M$. For small $M$, $r_{oc}$ can even approach $1$. This phenomenon actually underscores the rationale behind introducing the VR-based design for XL-MIMO. Specifically, as the number of antennas increases, the physical dimensions of the array also grow, and the portion of the array that receives marginal power increases. In this context, it is inefficient to use the whole array to compute the symbol detectors, since the majority of the power concentrates in a small portion of the array. Conversely, when $M$ is small, all sub-arrays may receive non-negligible power and therefore $r_{oc}$ approaches one. It can be seen from Fig. \ref{figure_multi_2} that for an XL-MIMO array with $M=10^4$, $25\%$ of the antennas receive $80\%$ of the total power (i.e., $\varpi=0.8$), which can be harnessed to substantially curtail the computational complexity in VR-based detection.

Fig. \ref{figure_multi_3} reveals the performance loss caused by the reduction of the complexity. As can be seen, when using ZF detection, the performance loss is small for moderate values of $\varpi $, whereas performance deteriorates when $\varpi$ is small. This is because as shown in Fig. \ref{figure_multi_2}, when $M=10^4$, less than $5\%$ of the antennas contribute $10\%$ of the power, $20\%$ of the antennas contribute $70\%$ of the power, and $70\%$ of the antennas contribute $99\%$ of the power. As a result, for a moderate value of $\varpi$, it is feasible to identify sub-arrays that receive the dominant power within the VR while encompassing a small number of antennas, which yields good performance at low  complexity. In other words, a moderate value of $\varpi$ can realize a good trade-off between  performance and  complexity. Besides, it can be observed that when $\varpi$ is small, the data rate degradation is severer under smaller $M$. This is because smaller arrays exhibit weaker spatial non-stationarity effects, and therefore more sub-arrays have to be used to receive sufficient power and attain satisfactory performance. Furthermore, for MRC, the VR-based design may outperform the WA-based design. The reason is that for VR-based MRC, the influence of multi-user interference diminishes when the VRs of users have less overlap.


\begin{figure}[t]
	\setlength{\abovecaptionskip}{0pt}
	\centering
	\includegraphics[width=3.4in]{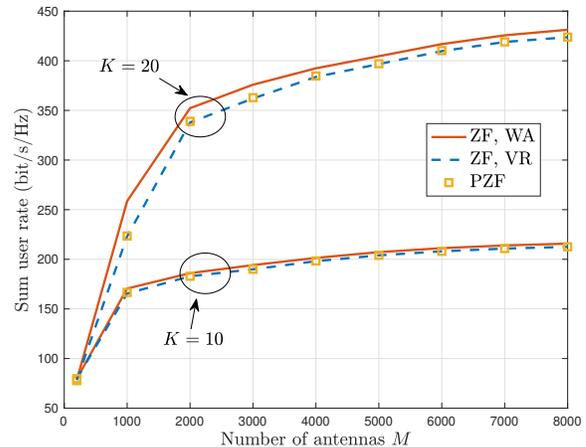}
	\caption{ Comparison of VR-based and PZF-based design, $\varpi=0.8$, $\hat{s}_{\mathrm{ovp}}=0.6$. }
	\label{figure_multi_4}
\end{figure}
\begin{figure}[t]
	\setlength{\abovecaptionskip}{0pt}
	\centering
	\includegraphics[width=3.4in]{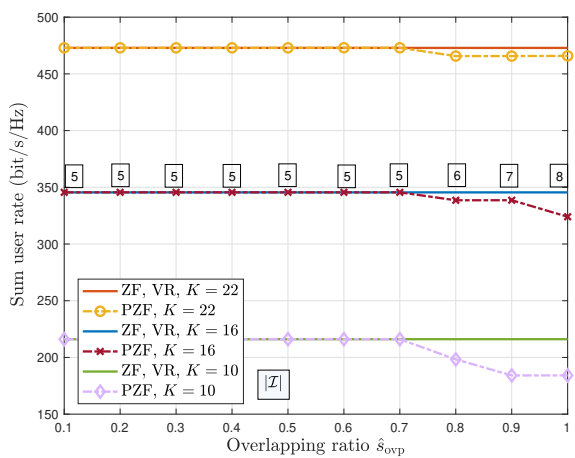}
	\caption{Sum user rate versus the threshold of VR overlapping ratio $\hat{s}_{\mathrm{ovp}}$.}
	\label{figure_multi_5}
\end{figure}

In Figs. \ref{figure_multi_4} and \ref{figure_multi_5}, we investigate the performance of the proposed user partitioning-based PZF algorithm. Fig. \ref{figure_multi_4} shows  that the proposed PZF detector can achieve almost the same performance as the VR-based detectors for any value of $M$ and for different values of $K$. This is because for the VR-based detectors,  users primarily suffer from the interference from other users having VRs with large overlap. Therefore, it is reasonable to partition users into different groups based on VR information, and then eliminate only the intra-group interference. Meanwhile, for the PZF detectors, a part of the channel degrees-of-freedom (DoFs) are used for interference nulling while the remaining DoFs are used to enhance the desired signal, which is beneficial for performance improvement.

The PZF detectors exploit user partitioning which is determined by VR overlap threshold $\hat{s}_{\mathrm{ovp}}$, as specified in Algorithm \ref{algorithm2}. In Fig. \ref{figure_multi_5}, we observe that the PZF algorithm only causes a performance loss for large $\hat{s}_{\mathrm{ovp}}$. This is because as $\hat{s}_{\mathrm{ovp}}$ increases, the criterion for establishing an edge between two vertices becomes more stringent, and thus the size of the independent set $|\mathcal{I}|$ could grow.  In other words, for large $\hat{s}_{\mathrm{ovp}}$,  users significantly interfering with each other may not be connected by edges, and they could be partitioned into different groups, and therefore the dominant interference may not be eliminated clearly by the intra-group PZF algorithm. With proper user partitioning (moderate $\hat{s}_{\mathrm{ovp}}$), the PZF algorithm can effectively address predominant interference and therefore cause negligible  performance loss at low complexity.

\begin{figure}[t]
	\setlength{\abovecaptionskip}{0pt}
	\centering
	\includegraphics[width=3.4in]{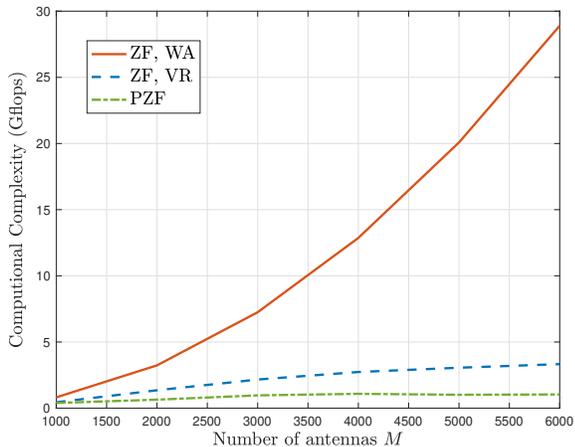}
	\caption{ Comparison of the computational complexity. }
	\label{figure_multi_6}
\end{figure}
Fig. \ref{figure_multi_6} shows the computational complexity associated with WA-based ZF, VR-based ZF, and user partition-based PZF detection. In accordance with Table \ref{tab1}, the complexity is computed based on the actual VR detection and user partitioning results utilizing MATLAB. As can be seen,  the complexity of the conventional WA-based design has a polynomial growth rate, which is not favorable considering the large number of antennas typically for XL-MIMO systems. However, by effectively exploiting the near-field spatial non-stationarities, the two proposed algorithms achieve much lower complexities.  VR-based ZF detection has a complexity that increases sub-linearly with the number of antennas, and the complexity of user partitioning-based PZF practically saturates for large $M$. The rationale behind this trend is twofold. On the one hand, the growth rate of the number of antennas within  VRs is considerably lower than $M$. As shown in Fig. \ref{figure_multi_2}, given $\varpi=0.7$, as $M$ increases from $10^3$ to $10^4$, the average number of antennas within VRs only increases from $600$ to $2000$. On the other hand, for large $M$, the physical dimensions of the array expand, and therefore, the VRs of different users become more separated on average. Accordingly, users can be divided into more groups, and then the computational complexity needed for PZF to eliminate the intra-group interference reduces.

\begin{figure}[t]
	\setlength{\abovecaptionskip}{0pt}
	\centering
	\includegraphics[width=3.4in]{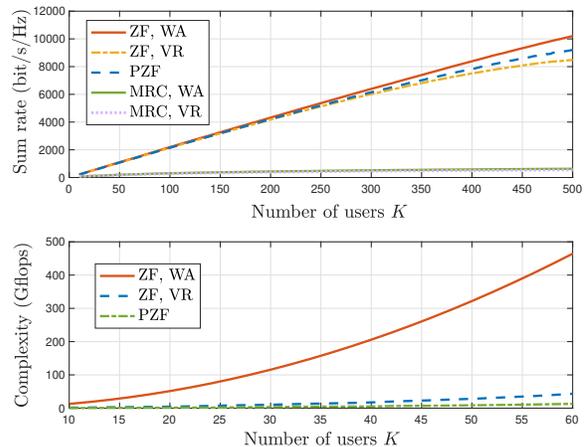}
	\caption{Sum user rate versus the number of users, $\varpi=0.8$.}
	\label{figure_multi_7}
\end{figure}
 Finally, Fig. \ref{figure_multi_7} illustrates the motivation and benefits of employing XL-MIMO. It can be seen that XL-MIMO has the capability to support a large number of users with extremely high throughput. As the number of users $K$ grows, small gaps emerge between the proposed low-complexity designs and the WA-based design. This is because the property of the multi-user interference becomes more and more complicated as $K$ increases. Nevertheless, the gaps can be effectively narrowed by increasing the value of $\varpi$. Meanwhile, it is shown that as $K$ increases, the proposed algorithms  exhibit significantly lower computational complexity. Therefore, the proposed VR-based and user partitioning-based PZF algorithms can effectively support extremely-high capacities in XL-MIMO systems at comparatively low complexity. Fig. \ref{figure_multi_7} also confirms the effectiveness and necessity of exploiting spatial non-stationarities when using the XL-MIMO.


\section{Conclusion}\label{section6}
In this work, we investigated the performance characteristics of XL-MIMO systems based on the EM channel model with near-field spatial non-stationarities. We derived an explicit expression for the SNR for the single-user scenario, which provided useful insights into the impact of the discrete aperture and polarization mismatch. We also complemented the classical  Fraunhofer distance under the proposed EM channel model. Building upon the near-field characteristics, we introduced a novel low-complexity linear detector based on VR information.  We also proposed a user partitioning algorithm grounded in graph theory, based on which PZF was used to further reduce the computational complexity. Simulation results validated the effectiveness of the proposed two low-complexity algorithms.

\begin{appendices}

\section{}\label{App_1}

	Define $\epsilon_x = \frac{\Delta_x}{r_{k,o}}$ and $\epsilon_y = \frac{\Delta_y}{r_{k,o}}$, where $r_{k,o}=\|\mathbf{u}_k\|$ is the distance between user $k$ and the origin. Substituting $\xi_{k,m_x, m_y}$ in (\ref{snr_sum}) with (\ref{pathloss3}), we have
\begin{align}
	&\mathrm{SNR}_k=\frac{p}{\sigma^2}\sum_{m_y\in\mathcal{M}_y} \sum_{m_x \in\mathcal{M}_x } \nonumber\\
	&\frac{A}{4 \pi} \frac{u_{k,z}\left(\left(m_x \Delta_x-u_{k,x}\right)^2+u_{k,z}^2\right)}{\left\{\left(m_x \Delta_x-u_{k,x}\right)^2+\left(m_y \Delta_y-u_{k,y}\right)^2+u_{k,z}^2\right\}^{\frac{5}{2}}}\\
	&=\frac{p}{\sigma^2}\sum_{m_y \in\mathcal{M}_y} \sum_{m_x \in\mathcal{M}_x }\nonumber\\
	& \frac{\eta \epsilon_x \epsilon_y}{4 \pi} \frac{\bar{u}_{k, z}\left(\left(m_x \epsilon_x-\bar{u}_{k, x}\right)^2+\bar{u}_{k, z}^2\right)}{\left\{\left(m_x \epsilon_x-\bar{u}_{k, x}\right)^2+\left(m_y \epsilon_y-\bar{u}_{k, y}\right)^2+\bar{u}_{k, z}^2\right\}^{\frac{5}{2}}},
\end{align}
where $\bar{u}_{k, c}=\frac{u_{k,c}}{r_{k,o}}$, $c\in\{x,y,z\}$.
Since $r_{k,o}\gg \Delta_x, \Delta_y$, we have $\epsilon_x, \epsilon_y \ll 1$. Then, we obtain (\ref{scsdcsc}) shown at the bottom of the page,
\begin{figure*}[b]
	\hrule
\begin{align}\label{scsdcsc}
	\begin{aligned}
		\mathrm{SNR}_k &\stackrel{(a)}{=} \frac{p}{\sigma^2}\frac{\eta  \epsilon_x \epsilon_y}{4 \pi} \sum_{\hat{y}=\frac{-\left(M_y-1\right) \epsilon_y}{2}}^{\frac{\left(M_y-1\right) \epsilon_y}{2}} \sum_{\hat{x}=\frac{-\left(M_x-1\right) \epsilon_x}{2}}^{\frac{\left(M_x-1\right) \epsilon_x}{2}} \frac{\bar{u}_{k,z}\left(\left(\hat{x}-\bar{u}_{k,x}\right)^2+\bar{u}_{k,z}^2\right)}{ \left\{\left(\hat{x}-\bar{u}_{k, x}\right)^2+\left(\hat{y}-\bar{u}_{k, y}\right)^2+\bar{u}_{k, z}^2\right\}^{\frac{5}{2}}}\\
		&\stackrel{(b)}{\approx} \frac{p}{\sigma^2} \frac{\eta \epsilon_x \epsilon_y}{4 \pi} \frac{1}{\epsilon_x \epsilon_y}
		\int_{\frac{-\left(M_y-1\right) \epsilon_y}{2} -\frac{\epsilon_y}{2}}^{\frac{\left(M_y-1\right) \epsilon_y}{2} + \frac{\epsilon_y}{2}}
		\int_{\frac{-\left(M_x-1\right) \epsilon_x}{2} -\frac{\epsilon_x}{2} }^{\frac{\left(M_x-1\right) \epsilon_x}{2} +\frac{\epsilon_x}{2} } \frac{\bar{u}_{k,z}\left(\left(\hat{x}-\bar{u}_{k,x}\right)^2+\bar{u}_{k,z}^2\right)}{\left\{\left(\hat{x}-\bar{u}_{k, x}\right)^2+\left(\hat{y}-\bar{u}_{k, y}\right)^2+\bar{u}_{k, z}^2\right\}^{\frac{5}{2}}} d \hat{x} d \hat{y}\\
		&\stackrel{(c)}{=} \frac{p}{\sigma^2} \frac{\eta \bar{u}_{k,z}}{4 \pi}  \int_{\frac{-M_x \epsilon_x}{2}-\bar{u}_{k,x}}^{\frac{M_x \epsilon_x}{2}-\bar{u}_{k,x}} \int_{\frac{-M_y \epsilon_y}{2}-\bar{u}_{k,y}}^{\frac{M_y \epsilon_y}{2}-\bar{u}_{k,y}} \frac{\left(x^2+\bar{u}_{k,z}^2\right)}{\left\{x^2+y^2+\bar{u}_{k,z}^2\right\}^{\frac{5}{2}}} d y d x,
	\end{aligned}
\end{align}
\end{figure*}
where the change of variables $\hat{x}=m_x \epsilon_x$ and $\hat{y}=m_y \epsilon_y$ is applied in $(a)$. In $(b)$, since $\epsilon_x, \epsilon_y\ll 1$, all variables within domain $[\hat{x}\pm \frac{\epsilon_x}{2}] \times [\hat{y}\pm \frac{\epsilon_y}{2}]$ of area $\epsilon_x\epsilon_y$ approximately yield the same objective function value as the center point $ ( \hat{x}, \hat{y} )$. Then, the double sum is approximated by a double integral divided by $\frac{1}{\epsilon_x \epsilon_y}$. The change of variables $ x=\hat{x}-\bar{u}_{k,x}$ and $ y=\hat{y}-\bar{u}_{k,y} $ are used in $(c)$. The proof can be completed by first solving  the integral with respect to $y$ based on \cite[2.271.6]{gradshteyn2014table}  and then solving the integral with respect to $x$ using \cite[(2.271.5)]{gradshteyn2014table} and \cite[(2.284)]{gradshteyn2014table}.

\section{}\label{App_2}
For the case of $M_y=1$, we obtain (\ref{sdsdfs}), shown at the bottom of the page.
\begin{figure*}[b]
	\hrule
\begin{align}\label{sdsdfs}
	\begin{aligned}
		\mathrm{SNR}_k&\!=\!\frac{p}{\sigma^2}\frac{\eta \epsilon_x\epsilon_y}{4 \pi} \sum\nolimits_{m_x=\frac{-\left(M_x-1\right)}{2}}^{\frac{\left(M_x-1\right)}{2}} \frac{\bar{u}_{k,z}\left(\left(m_x \epsilon_x-\bar{u}_{k,x}\right)^2+\bar{u}_{k,z}^2\right)}{ \left\{\left(m_x\epsilon_x-\bar{u}_{k, x}\right)^2+\bar{u}_{k, y}^2+\bar{u}_{k, z}^2\right\}^{\frac{5}{2}}}
		\!\approx\! \frac{p}{\sigma^2}\frac{\eta \epsilon_x\epsilon_y}{4 \pi} \frac{1}{\epsilon_x} \int_{\frac{-M_x\epsilon_x}{2}}^{\frac{M_x\epsilon_x}{2}} \frac{\bar{u}_{k,z}\left(\left(\hat{x}-\bar{u}_{k,x}\right)^2+\bar{u}_{k,z}^2\right)}{\left\{\left(\hat{x}-\bar{u}_{k,x}\right)^2+\bar{u}_{k,y}^2+\bar{u}_{k,z}^2\right\}^{\frac{5}{2}}} d \hat{x}\\
		&=\frac{p}{\sigma^2}\frac{\eta \epsilon_y}{4 \pi}  \bar{u}_{k,z} \int_{\frac{-M_x \epsilon_x}{2}-\bar{u}_{k,x} }^{\frac{M_x\epsilon_x}{2}-\bar{u}_{k,x}} \frac{1}{\left\{x^2+\bar{u}_{k,y}^2+\bar{u}_{k,z}^2\right\}^{\frac{3}{2}}}-\frac{\bar{u}_{k,y}^2}{\left\{x^2+\bar{u}_{k,y}^2+\bar{u}_{k,z}^2\right\}^{\frac{5}{2}}} d x.
	\end{aligned}
\end{align}
\end{figure*}
Utilizing \cite[(2.271.5)]{gradshteyn2014table} and \cite[(2.271.6)]{gradshteyn2014table}, we can complete the proof after some algebraic operations.

\end{appendices}

\bibliographystyle{IEEEtran}
\bibliography{myref.bib}

\begin{IEEEbiography}[{\includegraphics[width=1in,clip,keepaspectratio]{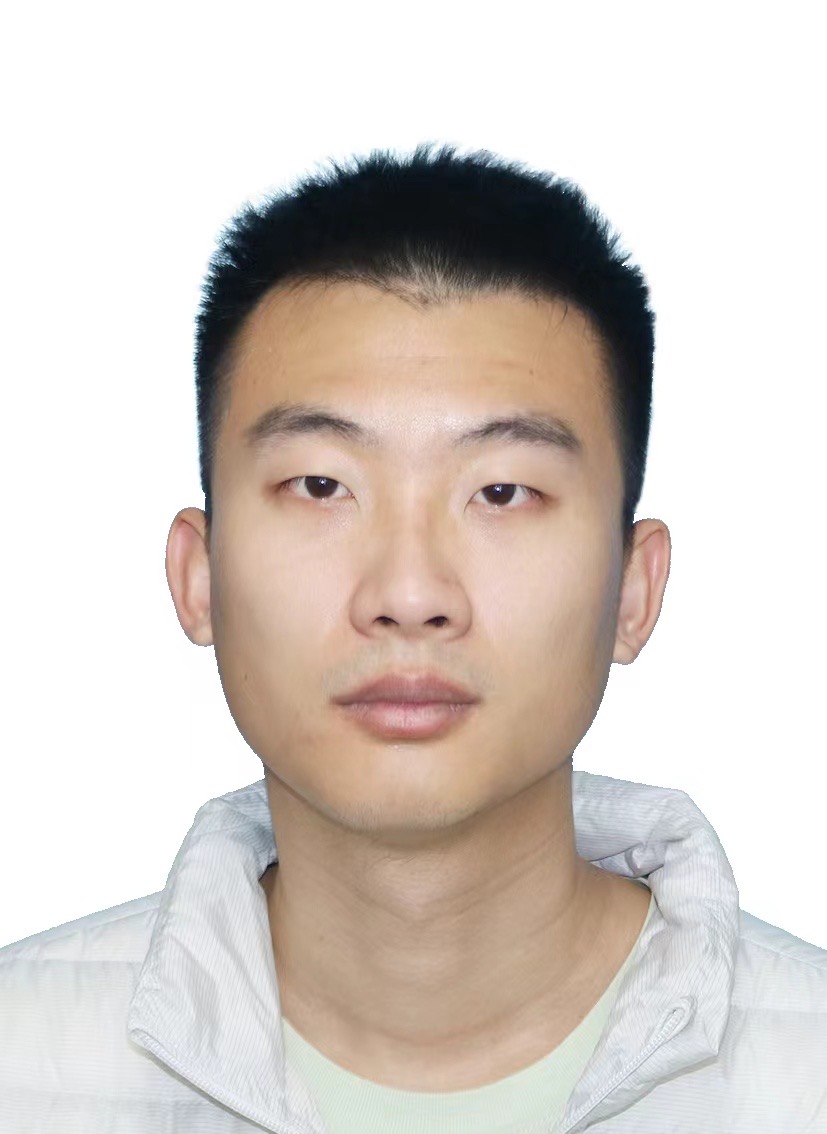}}]	
	{Kangda Zhi} received the B.Eng degree from the School of Communication and Information Engineering, Shanghai University (SHU), Shanghai, China, in 2017, the M.Eng degree from School of Information Science and Technology, University of Science and Technology of China (USTC), Hefei, China, in 2020, and the Ph.D. degree from the School of Electronic Engineering and Computer Science, Queen Mary University of London, U.K., in 2023. His research interests include Reconfigurable Intelligent Surface (RIS), massive MIMO, and near-field communications. He received the Exemplary Reviewer Certificate of the IEEE WIRELESS COMMUNICATIONS LETTERS in 2021 and 2022.
\end{IEEEbiography} 

\begin{IEEEbiography}[{\includegraphics[width=1in,clip,keepaspectratio]{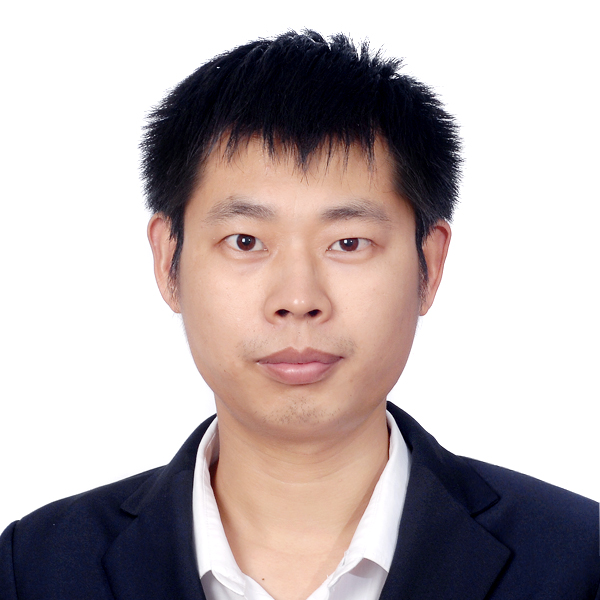}}]	
	{Cunhua Pan}
	received the B.S. and Ph.D. degrees from the School of Information Science and Engineering, Southeast University, Nanjing, China, in 2010 and 2015, respectively. From 2015 to 2016, he was a Research Associate at the University of Kent, U.K. He held a post-doctoral position at Queen Mary University of London, U.K., from 2016 and 2019.From 2019 to 2021, he was a Lecturer in the same university. From 2021, he is a full professor in Southeast University. 
	
	His research interests mainly include  reconfigurable intelligent surfaces (RIS), intelligent reflection surface (IRS), ultra-reliable low latency communication (URLLC) , machine learning, UAV, Internet of Things, and mobile edge computing. He has published over 120 IEEE journal papers. He is currently an Editor of IEEE Transactions on Vehicular Technology, IEEE Wireless Communication Letters, IEEE Communications Letters and IEEE ACCESS. He serves as the guest editor for IEEE Journal on Selected Areas in Communications on the special issue on xURLLC in 6G: Next Generation Ultra-Reliable and Low-Latency Communications. He also serves as a leading guest editor of IEEE Journal of Selected Topics in Signal Processing (JSTSP)  Special Issue on Advanced Signal Processing for Reconfigurable Intelligent Surface-aided 6G Networks, leading guest editor of IEEE Vehicular Technology Magazine on the special issue on Backscatter and Reconfigurable Intelligent Surface Empowered Wireless Communications in 6G, leading guest editor of IEEE Open Journal of Vehicular Technology on the special issue of Reconfigurable Intelligent Surface Empowered Wireless Communications in 6G and Beyond, and leading guest editor of IEEE ACCESS Special Issue on Reconfigurable Intelligent Surface Aided Communications for 6G and Beyond. He is Workshop organizer in IEEE ICCC 2021 on the topic of Reconfigurable Intelligent Surfaces for Next Generation Wireless Communications (RIS for 6G Networks), and workshop organizer in IEEE Globecom 2021 on the topic of Reconfigurable Intelligent Surfaces for future wireless communications. He is currently the Workshops and Symposia officer for Reconfigurable Intelligent Surfaces Emerging Technology Initiative. He is workshop chair for IEEE WCNC 2024, and TPC co-chair for IEEE ICCT 2022. He serves as a TPC member for numerous conferences, such as ICC and GLOBECOM, and the Student Travel Grant Chair for ICC 2019.  He received the  IEEE ComSoc Leonard G. Abraham Prize in 2022, IEEE ComSoc Asia-Pacific Outstanding Young Researcher Award, 2022.
	
\end{IEEEbiography} 

\begin{IEEEbiography}[{\includegraphics[width=1in,clip,keepaspectratio]{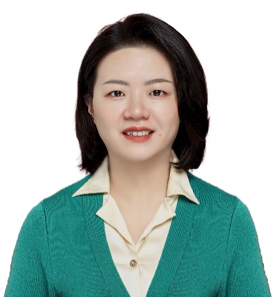}}]	
	{Hong Ren} received the B.S. degree in electrical engineering from Southwest Jiaotong University, Chengdu, China, in 2011, and the M.S. and Ph.D. degrees in electrical engineering from Southeast University, Nanjing, China, in 2014 and 2018, respectively. From 2016 to 2018, she was a Visiting Student with the School of Electronics and Computer Science, University of Southampton, U.K. From 2018 to 2020, she was a Post-Doctoral Scholar with Queen Mary University of London, U.K. She is currently an associate professor with Southeast University. Her research interests lie in the areas of communication and signal processing, including ultra-low latency and high reliable communications, Massive MIMO and machine learning.
\end{IEEEbiography}

\begin{IEEEbiography}[{\includegraphics[width=1in,clip,keepaspectratio]{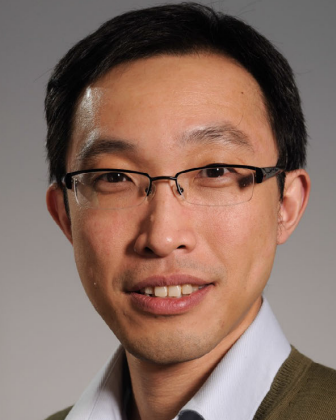}}]	
{Kok Keong Chai} received the B.Eng. (Hons.),
M.Sc., and Ph.D. degrees, in 1998, 1999, and
2007, respectively. He joined the School of
Electronic Engineering and Computer Science
(EECS), Queen Mary University of London
(QMUL), in August 2008. He is currently a Professor in the Internet of Things, the Queen Mary
Director of Joint Programme with the Beijing University of Posts and Telecommunications (BUPT),
and a Communication Systems Research Group
Member of QMUL. He has authored more than 65 technical journals and
conference papers in his research areas. His current research interests include
sensing and prediction in distributed smart grid networks, smart energy
charging schemes, applied blockchain technologies, dynamic resource management, wireless communications, and medium access control (MAC) for
M2M communications and networks.
\end{IEEEbiography}

\begin{IEEEbiography}[{\includegraphics[width=1in,clip,keepaspectratio]{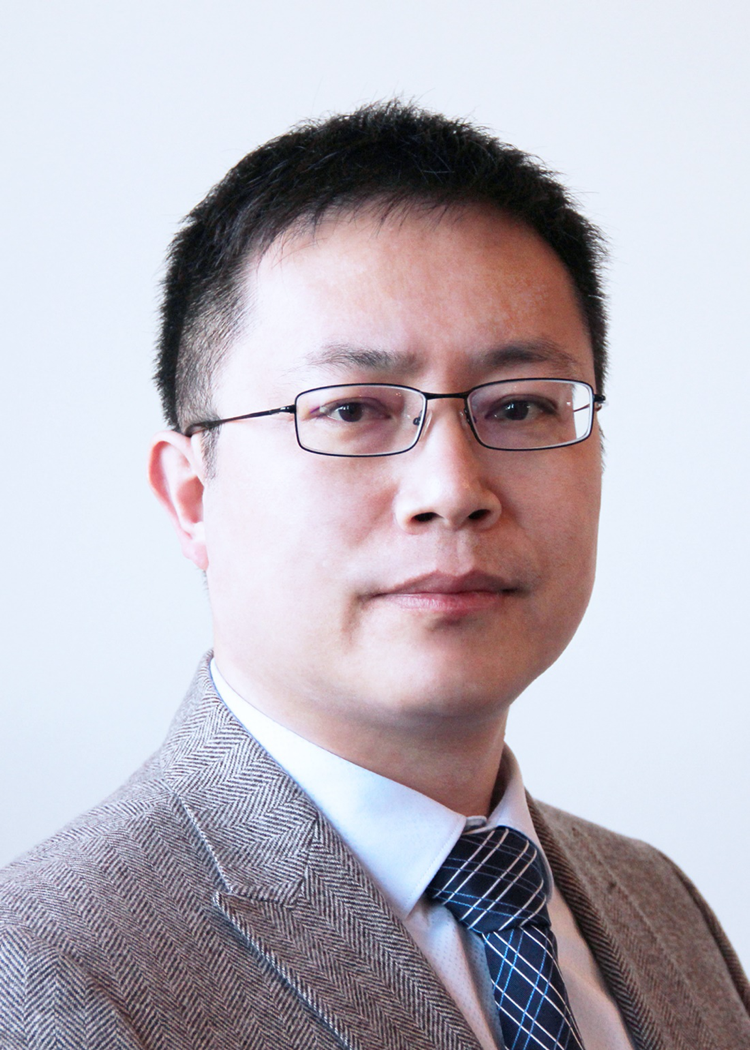}}]	
	{Cheng-Xiang Wang} (Fellow, IEEE) received the B.Sc. and M.Eng. degrees in communication and information systems from Shandong University, China, in 1997 and 2000, respectively, and the Ph.D. degree in wireless communications from Aalborg University, Denmark, in 2004.
	
	He was a Research Assistant with the Hamburg University of Technology, Hamburg, Germany, from 2000 to 2001, a Visiting Researcher with Siemens AG Mobile Phones, Munich, Germany, in 2004, and a Research Fellow with the University of Agder, Grimstad, Norway, from 2001 to 2005. He has been with Heriot-Watt University, Edinburgh, U.K., since 2005, where he was promoted to a Professor in 2011. In 2018, he joined Southeast University, Nanjing, China, as a Professor. He is also a part-time Professor with Purple Mountain Laboratories, Nanjing. He has authored 4 books, 3 book chapters, and 520 papers in refereed journals and conference proceedings, including 27 highly cited papers. He has also delivered 24 invited keynote speeches/talks and 16 tutorials in international conferences. His current research interests include wireless channel measurements and modeling, 6G wireless communication networks, and electromagnetic information theory.
	
	Dr. Wang is a Member of the Academia Europaea (The Academy of Europe), a Member of the European Academy of Sciences and Arts (EASA), a Fellow of the Royal Society of Edinburgh (FRSE), IEEE, IET and China Institute of Communications (CIC), an IEEE Communications Society Distinguished Lecturer in 2019 and 2020, a Highly-Cited Researcher recognized by Clarivate Analytics in 2017-2020. He is currently an Executive Editorial Committee Member of the IEEE TRANSACTIONS ON WIRELESS COMMUNICATIONS. He has served as an Editor for over ten international journals, including the IEEE TRANSACTIONS ON WIRELESS COMMUNICATIONS, from 2007 to 2009, the IEEE TRANSACTIONS ON VEHICULAR TECHNOLOGY, from 2011 to 2017, and the IEEE TRANSACTIONS ON COMMUNICATIONS, from 2015 to 2017. He was a Guest Editor of the IEEE JOURNAL ON SELECTED AREAS IN COMMUNICATIONS, Special Issue on Vehicular Communications and Networks (Lead Guest Editor), Special Issue on Spectrum and Energy Efﬁcient Design of Wireless Communication Networks, and Special Issue on Airborne Communication Networks. He was also a Guest Editor for the IEEE TRANSACTIONS ON BIG DATA, Special Issue on Wireless Big Data, and is a Guest Editor for the IEEE TRANSACTIONS ON COGNITIVE COMMUNICATIONS AND NETWORKING, Special Issue on Intelligent Resource Management for 5G and Beyond. He has served as a TPC Member, a TPC Chair, and a General Chair for more than 30 international conferences. He received 15 Best Paper Awards from IEEE GLOBECOM 2010, IEEE ICCT 2011, ITST 2012, IEEE VTC 2013 Spring, IWCMC 2015, IWCMC 2016, IEEE/CIC ICCC 2016, WPMC 2016, WOCC 2019, IWCMC 2020, WCSP 2020, CSPS2021, WCSP 2021, and IEEE/CIC ICCC 2022.
	
\end{IEEEbiography} 

\begin{IEEEbiography}[{\includegraphics[width=1in,clip,keepaspectratio]{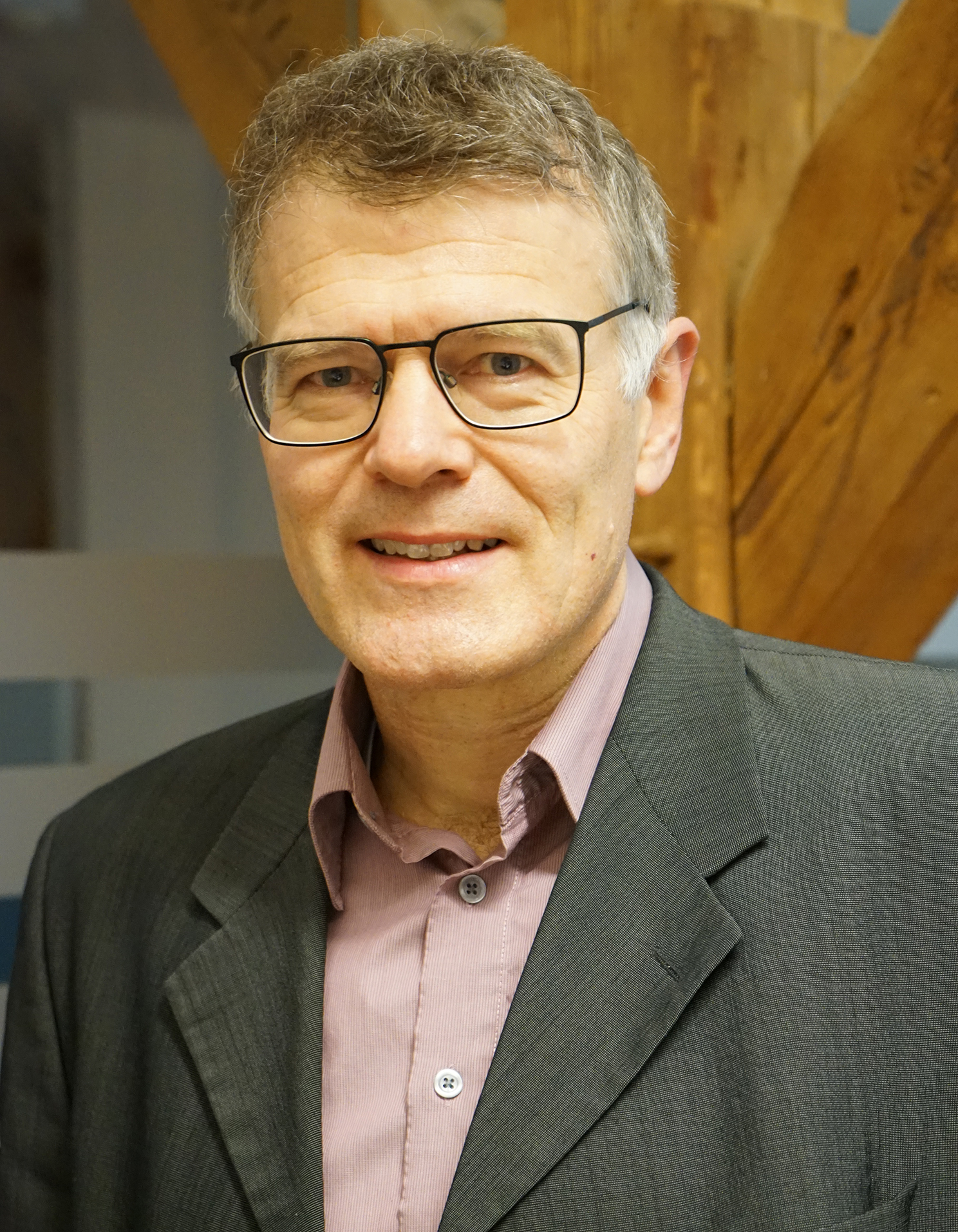}}]	
	{Robert Schober} (S'98, M'01, SM'08, F'10) received the Diplom (Univ.) and the Ph.D. degrees in electrical engineering from Friedrich-Alexander University of Erlangen-Nuremberg (FAU), Germany, in 1997 and 2000, respectively. From 2002 to 2011, he was a Professor and Canada Research Chair at the University of British Columbia (UBC), Vancouver, Canada. Since January 2012 he is an Alexander von Humboldt Professor and the Chair for Digital Communication at FAU. His research interests fall into the broad areas of Communication Theory, Wireless and Molecular Communications, and Statistical Signal Processing.
	
	Robert received several awards for his work including the 2002 Heinz Maier­ Leibnitz Award of the German Science Foundation (DFG), the 2004 Innovations Award of the Vodafone Foundation for Research in Mobile Communications, a 2006 UBC Killam Research Prize, a 2007 Wilhelm Friedrich Bessel Research Award of the Alexander von Humboldt Foundation, the 2008 Charles McDowell Award for Excellence in Research from UBC, a 2011 Alexander von Humboldt Professorship, a 2012 NSERC E.W.R. Stacie Fellowship, a 2017 Wireless Communications Recognition Award by the IEEE Wireless Communications Technical Committee, and the 2022 IEEE Vehicular Technology Society Stuart F. Meyer Memorial Award. Furthermore, he received numerous Best Paper Awards for his work including the 2022 ComSoc Stephen O. Rice Prize and the 2023 ComSoc Leonard G. Abraham Prize. Since 2017, he has been listed as a Highly Cited Researcher by the Web of Science. Robert is a Fellow of the Canadian Academy of Engineering, a Fellow of the Engineering Institute of Canada, and a Member of the German National Academy of Science and Engineering.
	
	He served as Editor-in-Chief of the IEEE Transactions on Communications, VP Publications of the IEEE Communication Society (ComSoc), ComSoc Member at Large, and ComSoc Treasurer. Currently, he serves as Senior Editor of the Proceedings of the IEEE and as ComSoc President-Elect.
\end{IEEEbiography} 

\begin{IEEEbiography}[{\includegraphics[width=1in,clip,keepaspectratio]{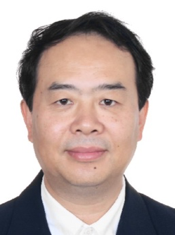}}]	
	{Xiao-Hu You} received his Master and Ph.D. Degrees from Southeast University, Nanjing, China, in Electrical Engineering in 1985 and 1988, respectively. Since 1990, he has been working with National Mobile Communications Research Laboratory at Southeast University, where he is currently professor and director of the Lab. He has contributed over 300 IEEE journal papers and 3 books in the areas of wireless communications. From 1999 to 2002, he was the Principal Expert of China C3G Project. From 2001-2006, he was the Principal Expert of China National 863 Beyond 3G FuTURE Project. From 2013 to 2019, he was the Principal Investigator of China National 863 5G Project. His current research interests include wireless networks, advanced signal processing and its applications. 
	
	Dr. You was selected as IEEE Fellow in 2011. He served as the General Chair for IEEE Wireless Communications and Networking Conference (WCNC) 2013, IEEE Vehicular Technology Conference (VTC) 2016 Spring, and IEEE International Conference on Communications (ICC) 2019. 
	
\end{IEEEbiography}

\end{document}